\documentclass[reprint,prd,twocolumn,superscriptaddress,showpacs,showkeys,floatfix,preprintnumbers]{revtex4-2}

\usepackage[english]{babel}
\usepackage[utf8]{inputenc}
\usepackage[T1]{fontenc}
\usepackage{amsmath,amssymb,braket,slashed,mathrsfs}
\usepackage{pifont}
\usepackage{hyperref}
\usepackage{color,hyperref}
\usepackage{cleveref}
\usepackage{amsmath}
\usepackage{graphicx}
\usepackage{booktabs}
\usepackage{multirow}
\usepackage{subfigure}
\usepackage{float}

\newcommand{\beq}{\begin{eqnarray}}
\newcommand{\eeq}{\end{eqnarray}}
\newcommand{\be}{\begin{equation}}
\newcommand{\ee}{\end{equation}}
\newcommand{\ba}{\begin{eqnarray}}
\newcommand{\ea}{\end{eqnarray}}

\newcommand{\bit}{\begin{itemize}}
\newcommand{\eit}{\end{itemize}}

\newcommand{\zzuphy}{School of Physics, Zhengzhou University, Zhengzhou 450001, China}
\newcommand{\innovation}{Collaborative Innovation Center of Quantum Matter, Beijing 100871, China}
\newcommand{\chep}{Center for High Energy Physics, Peking University, Beijing 100871, China}
\newcommand{\pkuphy}{School of Physics, Peking University, Beijing 100871,
China}

\newcommand{\CAS}{School of Physical Sciences, University of Chinese Academy of Sciences, Beijing 100049, People’s Republic of China}
\newcommand{\ITP}{CAS Key Laboratory of Theoretical Physics, Institute of Theoretical Physics, Chinese Academy of Sciences, Beijing 100190, China}
\newcommand{\IASHZ}{School of Fundamental Physics and Mathematical Sciences, Hangzhou Institute for Advanced Study, UCAS, Hangzhou 310024, China}
\newcommand{\ICTP}{International Centre for Theoretical Physics Asia-Pacific, Beijing/Hangzhou, China}
\newcommand{\CNIC}{Computer Network Information Center, Chinese Academy of Sciences, Beijing,100083, China}
\newcommand{\HISKP}{Helmholtz-Institut f\"ur Strahlen- und Kernphysik,  Universit\"at Bonn, 53115 Bonn, Germany}
\newcommand{\BNL}{Physics Department, Brookhaven National Laboratory, Upton, NY 11973, USA}

\begin{document}

\title{First lattice QCD calculation of $J/\psi$ semileptonic decay containing $D$ and $D_s$ particles }

\author{Yu Meng}
\email[Email: ]{yu_meng@zzu.edu.cn}
\affiliation{\zzuphy}
\author{Jin-Long Dang}
\affiliation{\pkuphy}
\author{Chuan Liu}
\email[Email: ]{liuchuan@pku.edu.cn}
\affiliation{\pkuphy}\affiliation{\chep}\affiliation{\innovation}
\author{Xin-Yu Tuo}
\affiliation{\BNL}
\author{Haobo Yan}
\affiliation{\pkuphy}\affiliation{\HISKP}
\author{Yi-Bo Yang}
\affiliation{\ITP}\affiliation{\CAS}\affiliation{\IASHZ}\affiliation{\ICTP}
\author{Ke-Long Zhang}
\email[Email: ]{klzhang@cnic.cn}
\affiliation{\CNIC}

\date{\today}

\begin{abstract}
We perform the first lattice calculation on the semileptonic decay of $J/\psi$ using the (2+1)-flavor Wilson-clover gauge ensembles generated by CLQCD collaboration. Three gauge ensembles with different lattice spacings, from 0.0519 fm to 0.1053 fm, and pion masses, $m_{\pi}\sim$ 300 MeV, are utilized.  After a naive continuum extrapolation using three lattice spacings, we obtain $\operatorname{Br}(J/\psi\rightarrow D_s e\nu_e)=1.90(6)(5)_{V_{cs}}\times 10^{-10}$ and $\operatorname{Br}(J/\psi\rightarrow D e\nu_e)=1.21(6)(9)_{V_{cd}}\times 10^{-11}$, where the first errors are statistical, and the second come from the uncertainties of CKM matrix element $V_{cs(d)}$. The ratios of the branching fractions between lepton $\mu$ and $e$ are also calculated as $R_{J/\psi}(D_s)=0.97002(8)$ and $R_{J/\psi}(D)=0.97423(15)$ after performing a continuum limit including only $a^2$ term. The ratios provide necessary theoretical support for the future experimental test of lepton flavor universality.

\end{abstract}

\maketitle

\section{Introduction}
The high-precision frontier of flavor physics is one of the most important goals in contemporary particle physics. It not only involves a precise test of the standard model but also closely relates to the search for new physics beyond the standard model. This type of test is typically categorized into two approaches. The first involves directly comparing with experimental results to identify any significant discrepancies, for example, the particle mass, leptonic decay constant, and the decay rate of a particular physical process. The second approach entails studying various flavor-changing weak decay processes to accurately extract the Cabibbo-Kobayashi-Maskawa(CKM) matrix elements, thereby assessing the unitarity of the CKM matrix elements in a row or column~\cite{FlavourLatticeAveragingGroupFLAG:2021npn}. In both cases, precise theoretical calculations and experimental measurements are essential and indispensable. Historically, however, the two are not always given at the same time, and the high-precision result that appears first often provides an important reference for the test of the other.

The discovery of the $J/\psi$ particle in 1974 has led to a revolution in modern particle physics and serves as a solid experimental foundation for establishing the standard model~\cite{Jpsi1974a,Jpsi1974b}. The main decay channels of $J/\psi$ particle are strong decay and electromagnetic decay, and these properties have been extensively studied for several decades both experimentally and theoretically. For the weak decay, the first estimated branching ratio of $J/\psi$ weak decay is at the order of $10^{-8}$~\cite{Sanchis-Lozano:1993vyw}, which is far from the experimental detection accuracy for a long time. 
With the BESIII experiment already accumulating more than 10 billion $J/\psi$ events~\cite{BESIII:2021cxx}, and even larger statistics for future Super Tau Charm Facility~\cite{Achasov:2023gey}, these weak decays will be marginally possible to be measured. Therefore, such rare decay has attracted much attention in recent years from the experiments~\cite{BES:2006mls,BESIII:2014pps,BESIII:2021mnd,BESIII:2023fqz} and various phenomenological studies, for example, Bauer-Stech-Wirbel (BSW) model~\cite{Dhir:2009rb}, the QCD sum rules (QCDSR)~\cite{Wang:2007ys}, the Bethe-Salpeter (BS) method~\cite{Wang:2016dkd}, covariant light-front quark model (CLFQM)~\cite{Sun:2023uyn}. However, these phenomenological calculations are rather precarious especially in the charm sector since the $J/\psi$ particle is compactly bound by the nonperturbative strong interaction. 
At the present stage, a genuine nonperturbative calculation can not only provide a model-independent comparison with previous phenomenological studies but also provide a precise and more reliable theoretical assist for future experiments. 

This work aims to present the first lattice calculation on $J/\psi$ semileptonic decay, and particularly determine the branching fraction $\operatorname{Br}(J/\psi\rightarrow D/D_s l \nu_l)$ with $l=e,\mu$. To that end, we have developed a method capable of simultaneously extracting multiple form factors, requiring only direct inputs of the lattice data. By constructing several appropriate scalar functions, the related form factors are calculated directly. Besides, we also calculate a large number of time separations between the initial and final particles in our simulation. Only when the ground states of initial and final particles dominate, a plateau independent of the time separation will emerge, thus effectively avoiding the mixing of excited states. To further control the discretization effect, three gauge ensembles with different lattice spacings are utilized and a continuum limit is finally performed. 

The rest of this paper is organized as follows. In Sec.~\ref{sec:method}, we
introduce the methodology utilized in this work for calculating the semileptonic decay rate. This section is divided into three parts: in Sec.~\ref{sec:decay_rate} a relationship between $J/\psi\rightarrow D/D_s l\nu_l$ decay width and relevant form factors are given; in section.~\ref{sec: form_factor}, the scalar function method is introduce to obtain the form factors on the lattice; in Sec.~\ref{sec:relation_M_E} the hadronic function is extracted from the lattice data. In Sec.~\ref{sec:result} we give details of the simulations and show the main results. This section is further divided into four parts: in Sec.~\ref{sec:mass_spectrun} the numerical values of $D_s$ and $D$ masses, together with the dispersion relation are presented; in Sec.~\ref{sec:form_factor_result} the numerical results of four form factors are presented; in section~\ref{sec:decay_width_result} the decay width is obtained and a continuum extrapolation under three lattice spacings is performed; 
in Sec.~\ref{sec:diff_decay_width_result} the differential decay width of $J/\psi\rightarrow D/D_s l\nu_l$ are presented for a better comparison with the future experiment; in section.~\ref{sec:discussion}, a relationship with traditional form factors is established. Finally, we conclude in Sec.~\ref{sec:conclude}.

\section{Methodology}\label{sec:method}
\subsection{Decay width of $J/\psi\rightarrow D/D_s l\nu_l$}\label{sec:decay_rate}
The amplitude of $J/\psi\rightarrow D/D_s l\nu_l$ in the lowest-order standard model is expressed by
\be\label{eq:amp}
i\mathcal{M}=-i\frac{G_F}{\sqrt{2}}V_{cs(d)}H_{\mu}(p,p')g_{\mu\nu}\bar{u}_{l}\gamma_{\nu}(1-\gamma_5)u_{\nu_l} 
\ee
where $l=e,\mu$. $p'$ and $p$ are the four-momenta of $J/\psi$ and $D/D_s$ particles, respectively. Here we introduce the $H_{\mu}(p,p')$ to denote the nonperturbative hadronic interaction between the initial and final states,
\be
H_{\mu}(p,p')\equiv \langle D/D_s(p)| J_{\mu}^{W}|J/\psi(\epsilon,p')\rangle
\ee
Such hadronic function can be generally decomposed by four variant form factors in the following way\cite{ZPC1990}
\beq\label{eq:FF}
H_{\mu}(p,p')&=&\epsilon_{\alpha}(p')H_{\mu\alpha}(p,p') \nonumber \\
H_{\mu\alpha}(p,p')&=&F_1(q^2)g_{\mu\alpha}+\frac{F_2(q^2)}{M m}p'_{\mu}p_{\alpha}+\frac{F_3(q^2)}{m^2}p_{\mu}p_{\alpha} \nonumber \\
&-& \frac{iF_{0}(q^2)}{M m}\epsilon_{\mu\alpha\rho\sigma}p'_{\rho}p_{\sigma}
\eeq
where $M$ is the $J/\psi$ mass and $m$ the $D/D_s$ mass. The $\epsilon_{\alpha}(p')$ is the polarization vector of $J/\psi$ particle. Compared to the previous parameterization\cite{ZPC1990}, the form factors in Eq.(\ref{eq:FF}) have been rescaled and have the same dimension.

A direct calculation on the decay width of $J/\psi \rightarrow D/D_s l \nu_l$ in the rest frame of $J/\psi$, by employing Eq.~(\ref{eq:amp}) and Eq.~(\ref{eq:FF}), leads to
\beq\label{eq:width}
\Gamma&=&\frac{1}{2 M}\int\frac{d^3\vec{p}}{(2\pi)^3 2E}\int\frac{d^3\vec{q_l}}{(2\pi)^32E_l}\int\frac{d^3\vec{q}_{\nu}}{(2\pi)^32E_{\nu_l}} \nonumber \\
&\times&(2\pi)^4\delta^4(p'-p-q_l-q_{\nu})\frac{1}{3}|\mathcal{M}|^2  \nonumber \\
&=&\frac{G_F^2V_{cs(d)}^2}{12M^2}\frac{1}{32\pi^3}\int_{m_l^2}^{(M-m)^2} dq^2 \times \big{[} c_0(E_l^+-E_l^-) \nonumber \\
&+&\frac{c_1}{2}((E_l^+)^2-(E_l^-)^2)+\frac{c_2}{3}((E_l^+)^3-(E_l^-)^3)\big{]}  \nonumber \\
\eeq
where $(q^2,E_l)$ are two independent Dalitz variables. The $q^2$ is the momentum transfer squared, $E$ the energy of $D/D_s$ particle, $m_l$ the lepton's mass, $q_l$ the lepton's four-momentum, $E_l$ the lepton's energy, $q_{\nu_l}$ the neutrino's four-momentum, and $E_{\nu_l}$ the neutrino's energy. Note that the $E$ is related to $q^2$ by $ E= (M^2+m^2-q^2)/(2M)$. The coefficients $c_i(i=0,1,2)$ are combinations of $M,m,m_l,E,F_i(i=0,1,2,3)$ and are therefore $q^2$-dependent. The expressions of $c_i$ are summarized in the Appendix{~\ref{sec:c_coeff}}. 

The phase space boundary in the $(q^2,E_l)$-plane is constrained by\cite{ZPC1990}
\beq
E_l^{\pm}&=&\frac{1}{2M}\Big{[} q^2+m_l^2-\frac{1}{2q^2}\big{(}(q^2-M^2+m^2)(q^2+m_l^2) \nonumber \\
&\mp& 2M|\vec{p}|(q^2-m_l^2)\big{)}\Big{]} 
\eeq
where $\vec{p}$ is the momentum of $D/D_s$ particle and $2M|\vec{p}|=\sqrt{(M^2-m^2)^2+q^4-2q^2(M^2+m^2)}$.

\subsection{Extraction of the form factor on the lattice}\label{sec: form_factor}
In this section, we proceed to extract the form factor $F_i(i=0,1,2,3)$ using the scalar function method, which has been widely applied to various physical processes in recent years~\cite{Feng:2019geu,Feng:2020zdc,Tuo:2021ewr,Meng:2021ecs,Zou:2021mgf,Fu:2022fgh,Tuo:2022hft,Christ:2022rho,Meng:2023bjc,Meng:2024gpd} and achieved great successes.  

We start our discussion with a Euclidean hadronic function in the infinite volume
\be
H_{\mu\nu}(\vec{x},t)=\langle 0| \phi_h(\vec{x},t)J_{\mu}^{W}(0)| J/\psi_{\nu}(\epsilon,p') \rangle, t>0
\ee
where $\phi_h$ is the interpolating operator of $D/D_s$ particle with $h=D$ or $h=D_s$, respectively. $J_\mu^{W}=\bar{s}\gamma_{\mu}(1-\gamma_5)c$ for $h=D_s$ and $J_\mu^{W}=\bar{l}\gamma_{\mu}(1-\gamma_5)c$ for $h=D$, here $l,s,c$ are the light, strange, and charm quarks. The $|J/\psi_{\nu}(\epsilon,p')\rangle$ is the $J/\psi$ state with specific polarization direction $\nu$ and momentum $p'=(iM,\vec{0})$. At large time $t$, the hadronic function is saturated by the single $h=D$ or $D_s$ state
\beq
&&H_{\mu\nu}(x)\doteq H_{\mu\nu}^{h}(x)=\int \frac{d^3\vec{p}}{(2\pi)^3}\frac{1}{2E_{h}}e^{-E_{h}t+i\vec{p}\cdot \vec{x}} \nonumber \\
&&\times \langle 0|\phi_h(0)|\phi_h(\vec{p})\rangle \langle \phi_h(\vec{p})|J_{\mu}^W(0)|J/\psi_{\nu}(p')\rangle
\eeq
Considering the following parameterizations
\beq\label{eq:F_param}
\langle 0|\phi_h(0)|\phi_h(\vec{p})\rangle &=&Z_{h} \nonumber \\
\langle\phi_h(\vec{p})|J_{\mu}^V(0)|J/\psi_{\nu}(\epsilon,p')\rangle&=&\frac{F_0(q^2)}{Mm}\epsilon_{\mu\nu\rho\sigma}p'_{\rho}p_{\sigma} \nonumber\\
\langle\phi_h(\vec{p})|J_{\mu}^A(0)|J/\psi_{\nu}(\epsilon,p')\rangle&=&-F_1(q^2)\delta_{\mu\nu}-\frac{F_2(q^2)}{M m}p'_{\mu}p_{\nu} \nonumber \\
&-&\frac{F_3(q^2)}{m^2}p_{\mu}p_{\nu} 
\eeq
where we have divided the weak current $J_{\mu}^{W}$ into two parts, i.e. $J_{\mu}^W=J_{\mu}^V-J_{\mu}^A$ for convenience. The four-momentum of the final state $\phi_h$ is denoted by $p=(iE_h,\vec{p})$ and the transfer momentum square is defined by $q^2=(M-E_h)^2-|\vec{p}|^2$ as $J/\psi$ is at rest. Then, the spatial Fourier transform of $H_{\mu\nu}(\vec{x},t)\equiv V_{\mu\nu}(\vec{x},t)-A_{\mu\nu}(\vec{x},t) $ yields
\beq
\tilde{V}_{\mu\nu}(\vec{p},t)&\doteq& \tilde{V}_{\mu\nu}^{h}(\vec{p},t)=\frac{F_0(q^2)}{Mm} \frac{Z_{h}}{2E_{h}}e^{-E_{h}t}\epsilon_{\mu\nu\rho\sigma}p'_{\rho}p_{\sigma}  \nonumber\\
\tilde{A}_{\mu\nu}(\vec{p},t)&\doteq & \tilde{A}_{\mu\nu}^{h}(\vec{p},t)=\frac{Z_{h}}{2E_{h}}e^{-E_{h}t}\big{(}-F_1(q^2)\delta_{\mu\nu}\nonumber \\ 
&-&\frac{F_2(q^2)}{M m}p'_{\mu}p_{\nu} -\frac{F_3(q^2)}{m^2}p_{\mu}p_{\nu} \big{)}   
\eeq
Similar to our previous studies~\cite{Meng:2021ecs,Meng:2024gpd}, the form factor $F_0$ can be accessed to a scalar function 
\beq
\mathcal{I}_0&\equiv &\frac{1}{M|\vec{p}|^2}\epsilon_{\mu\nu\rho \sigma}p'_{\rho}p_{\sigma}\tilde{V}_{\mu\nu}(\vec{p},t) 
\eeq
and it is obtained by
\beq\label{eq:V_eff}
F_0(q^2)&=&\frac{m E_h }{Z_h} e^{E_ht} \int d^3\vec{x}\frac{j_1(|\vec{p}||\vec{x}|)}{|\vec{p}||\vec{x}|}\epsilon_{\mu\nu \alpha 0}x_{\alpha}V_{\mu\nu}(\vec{x},t)  \nonumber \\
\eeq
where $j_n(x)$ are the spherical Bessel functions. To extract $F_i(i=1,2,3)$, we construct the following scalar functions
\beq
\mathcal{I}_1&\equiv& \delta_{\mu\nu}\tilde{A}_{\mu\nu}(\vec{p},t) \nonumber\\
\mathcal{I}_2&\equiv& \frac{E}{M}\frac{ p'_{\mu}p_{\nu}}{|\vec{p}|^2}\tilde{A}_{\mu\nu}(\vec{p},t)\nonumber\\
\mathcal{I}_3&\equiv& \frac{p_{\mu}p_{\nu}}{|\vec{p}|^2}\tilde{A}_{\mu\nu}(\vec{p},t)
\eeq
and it immediately leads to
\beq
F_1(q^2)&=&\frac{2E_h e^{E_ht}}{3m^2Z_h}\left[E_h^2I_{2}-E_h|\vec{p}|^2(I_{3}+I_{4})-m^2I_1-|\vec{p}|^2I_5 \right] \nonumber \\
F_2(q^2)&=&\frac{2E_h e^{E_ht}}{mZ_h}\left[E_hI_{2}-E_h^2I_{4}-E_hI_5-|\vec{p}|^2I_{3}\right] \nonumber \\
F_3(q^2)&=&\frac{2E_h e^{E_ht}}{3m^2Z_h}\big{[}E_h^2I_{2}+3m_{h}^2(E_hI_{4}+I_5)-m^2I_1 \nonumber\\
&-&|\vec{p}|^2(E_hI_{3}+E_hI_{4}+I_5)\big{]}
\eeq
where $I_i(i=1,2,3,4,5)$ are scalar functions that are directly related to the correlation functions $A_{\mu\nu}(\vec{x},t)$.  The expressions of these $I_i$ are listed in the Appendix{~\ref{sec:I_scalar}}. 

Note that both $V_{\mu\nu}$ and $A_{\mu\nu}$ denote the correlation function in the infinite volume. In real lattice simulations, we need to use the hadronic function $V_{\mu\nu}^{L}(\vec{x},t)$ and $A_{\mu\nu}^{L}(\vec{x},t)$ which are calculated on a finite-volume lattice to replace the infinite-volume $V_{\mu\nu}(\vec{x},t)$ and $A_{\mu\nu}(\vec{x},t)$. Such replacement only amounts for exponentially suppressed finite-volume effects as they are suppressed exponentially when $|\vec{x}|$ becomes large. One can introduce a spatial integral truncation $R=|\vec{x}|$ and examine at large $R$ whether the finite-volume effects are well under control or not, as we did in our previous study~\cite{Meng:2021ecs}. In our latter work on the $D_s^*$ radiative decay~\cite{Meng:2024gpd}, such inspections have also been performed, and the finite-volume effects are confirmed to be well-controlled (this part of the analysis was not included in the article). In this work, we also present the examination and the results are summarized in the Appendix~\ref{sec:FV_effect}. It is also observed that the finite-volume effects can be negligible. Therefore, we can empirically conclude that for the case where the intermediate state is a charm or heavier meson, we can make the substitution directly without any additional operations. However, when the intermediate state is a light meson, e.g. $\pi$ or $K$, the exponentially suppressed finite-volume effects may not be ignored. In that case, one can use, for example, the infinite volume reconstruction(IVR) method to deal with these problems~\cite{Tuo:2021ewr}.

\subsection{Hadronic function on the lattice}\label{sec:relation_M_E}
The hadronic function $H_{\mu\nu}(\vec{x},t)$ can be
extracted from a three-point function
\beq
C_{\mu\nu}^h(\vec{x},t;t_s)=\langle \mathcal{O}_{h}(t)J_{\mu}^{W}(0)\mathcal{O}_{J/\psi,\nu}^{\dagger}(-t_s) \rangle 
\eeq
where interpolating operators are chosen as $\mathcal{O}_{J/\psi,\nu}^{\dagger}=-\bar{c}\gamma_{\mu}c$, $\mathcal{O}_{h=D}=\bar{l}\gamma_5 c$, and $\mathcal{O}_{h=D_s}=\bar{s}\gamma_5 c$ . In this work, we only consider the
connected contribution. Then, it has the following contractions
\beq
&&C_{\mu\nu}^{D}(\vec{x},t;t_s) \nonumber \\
&=&\langle \gamma_{\nu}\gamma_5S_{c}^{\dagger}(t,-t_s)\gamma_5\gamma_5S_l(t,0)\gamma_{\mu}(1-\gamma_5)S_c(0,-t_s) \rangle \nonumber \\
&&C_{\mu\nu}^{D_s}(\vec{x},t;t_s) \nonumber \\
&=&\langle \gamma_{\nu}\gamma_5S_{c}^{\dagger}(t,-t_s)\gamma_5\gamma_5S_s(t,0)\gamma_{\mu}(1-\gamma_5)S_c(0,-t_s) \rangle \nonumber \\
\eeq
Then, the hadroinc function $H_{\mu\nu}(\vec{x},t)$ is determined
directly through
\beq
H_{\mu\nu}(\vec{x},t)=\frac{2M}{Z_M}e^{Mt}C_{\mu\nu}^{h}(\vec{x},t)
\eeq
In our calculations, $M$, $m$, $Z_M$, and $Z_{h}$ are extracted from the two-point function $C^{(2)}(\vec{p},t)=\sum\limits_{\vec{x}}\cos(\vec{p}\cdot \vec{x})\langle\mathcal{O}_{h}(\vec{x},t)\mathcal{O}_{h}^{\dagger}(0)\rangle $ by a single-state fit
\beq\label{eq:2pt}
C^{(2)}(\vec{p},t)=\frac{Z_{h}^2}{2E_{h}}\left( e^{-E_{h} t}+e^{-E_{h}(T-t)} \right)
\eeq
with $m_{h}=E_{h}(\vec{p}=0)$ the ground-state energy and $Z_h=\langle h|\mathcal{O}^{\dagger}_h|0 \rangle$ is the overlap amplitude for the ground state. The calculation of three-point function $C_{\mu\nu}^{h}(\vec{x},t)$ is similar to our previous studies~\cite{Meng:2021ecs,Meng:2024gpd}, so we won't go into these details in this paper.

\section{Simulations and results}\label{sec:result}
\begin{table}[!h]
\begin{ruledtabular}
\begin{tabular}{cccc}
\textrm{Ensemble} & C24P29 & F32P30 & H48P32 \\
\hline
$a(\textrm{fm})$ & 0.10530(18) &0.07746(18) & 0.05187(26) \\
$a\mu_s$ & -0.2400 & -0.2050 &-0.1700 \\
$a\mu_c$ &0.4479 &0.2079 & 0.0581 \\
$L^3\times T$ & $24^3\times 72$ & $32^3\times 96$ & $48^3\times 144$ \\
$N_{\textrm{cfg}}\times N_{\textrm{src}}$ & $450\times 72$ & $719\times 96$ & $100\times 72$ \\
$m_{\pi}(\textrm{MeV})$ & 292.7(1.2) & 303.2(1.3) & 317.2(0.9) \\
$t$ & 6-17 & 6-20& 8-30 \\
$Z_V$ & 0.79814(23)& 0.83548(12)& 0.86855(04) \\
$Z_A$ & 0.85442(85)& 0.88161(64)& 0.90113(36) \\
\end{tabular}
\end{ruledtabular}
\caption{
Parameters of gauge ensembles used in this work. From top to bottom, we list the ensemble name, the lattice spacing $a$,
the bare quark mass including the strange quark $a\mu_s$ and charm quark $a{\mu}_c$, the spatial and temporal lattice size $L$ and $T$,
the number of the measurements of the correlation function for each ensemble $N_{\textrm{cfg}}\times N_{\textrm{src}}$, the pion mass $m_{\pi}$, the range of the time separation $t$ between the initial hadron and the electromagnetic current, vector normalization constant $Z_V$, and axial vector normalization constant $Z_A$.
Here, $L$, $T$ and $t$ are given in lattice units.\label{table:cfg}}
\end{table}

We employ three (2+1)-flavor Wilson-clover gauge ensembles generated by the CLQCD collaboration with lattice spacings $a\approx  0.1053,0.0775,0.0519$ fm~\cite{CLQCD:2023sdb}, the parameters of which are shown in Table.~\ref{table:cfg}. The dynamical quarks of the ensembles use the tadpole-improved tree-level Symanzik gauge action and the tadpole-improved tree-level clover fermions. In recent years, plenty of studies have been carried out based on these configurations~\cite{Zhang2022, Liu2023, Liu2024, Liu2024a,Meng:2024gpd,Yan:2024yuq,Yan:2024gwp,Du:2024wtr}. The same setup has been utilized in our previous lattice study on $D_s^*$ radiative decay. For more details of the ensembles, we refer to Ref.~\cite{Meng:2024gpd}. Since the ensembles have similar volumes and pion masses in physical units and are therefore expected to provide a fully well-controlled continuous extrapolation. 

\subsection{Mass spectrum and dispersion relation}\label{sec:mass_spectrun}

\begin{figure*}[htbp]
\centering
\subfigure{
\centering
\includegraphics[width=8.5cm]{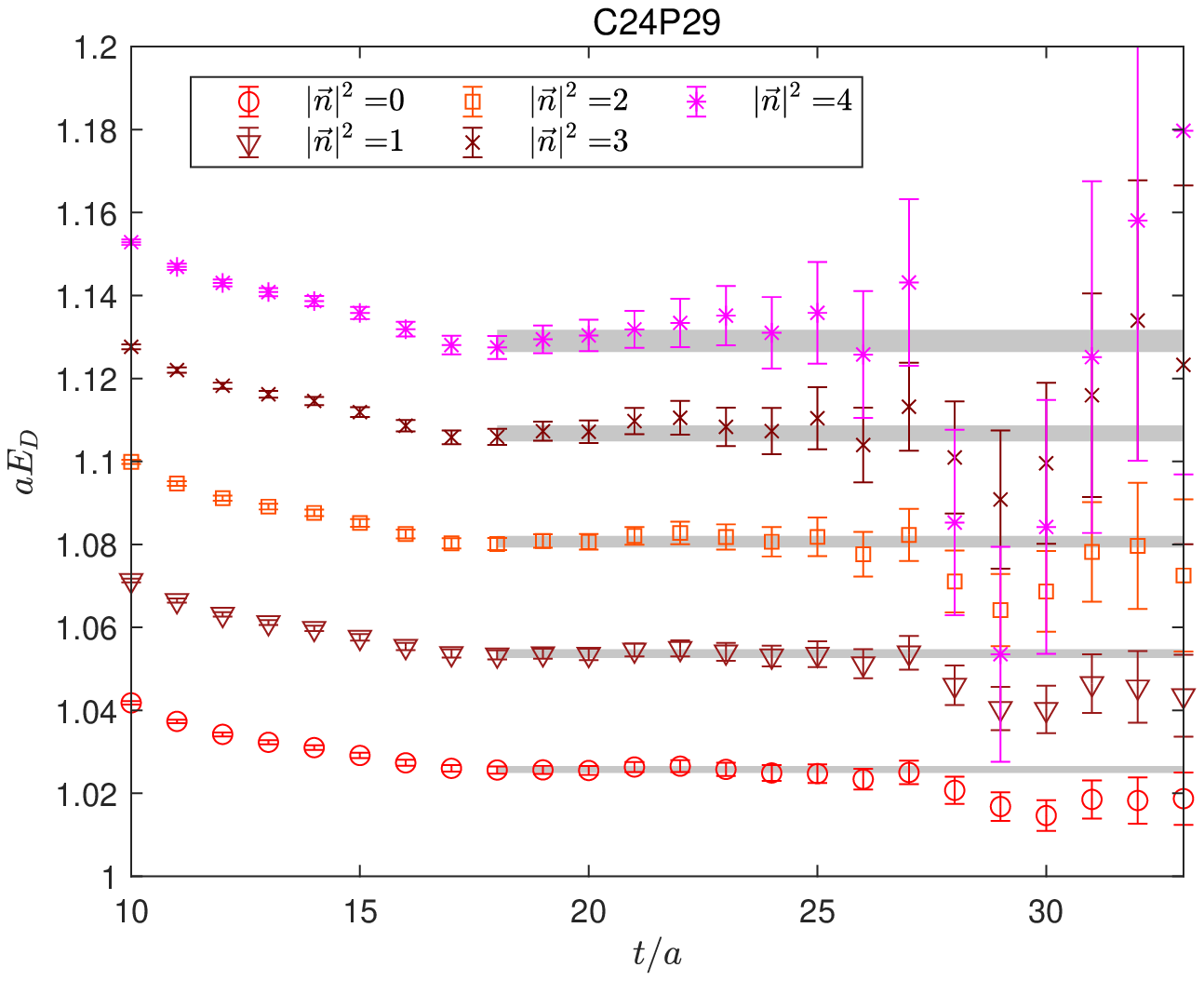}
}
\subfigure{
\centering
\includegraphics[width=8.5cm]{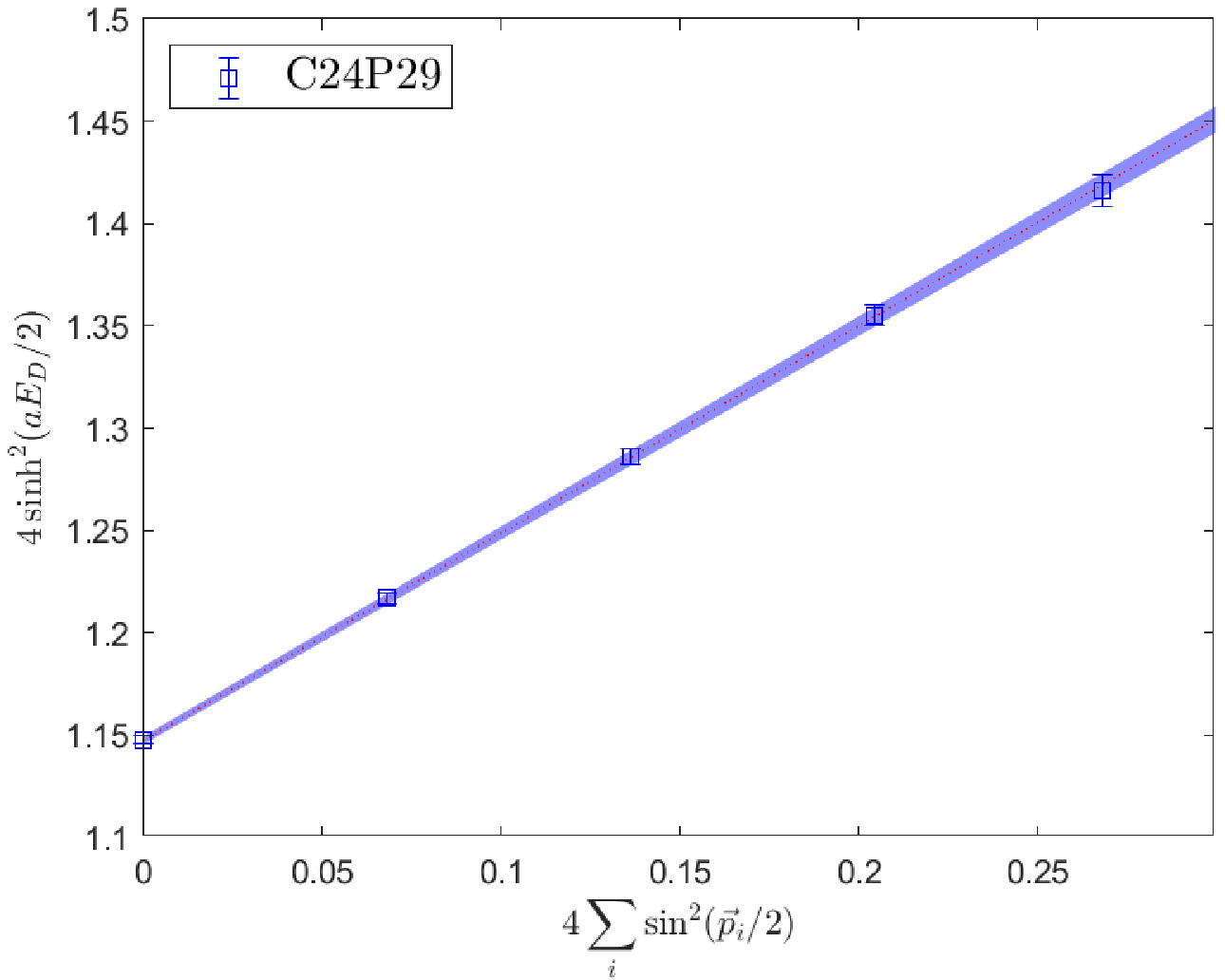}
}
\subfigure{
\centering
\includegraphics[width=8.5cm]{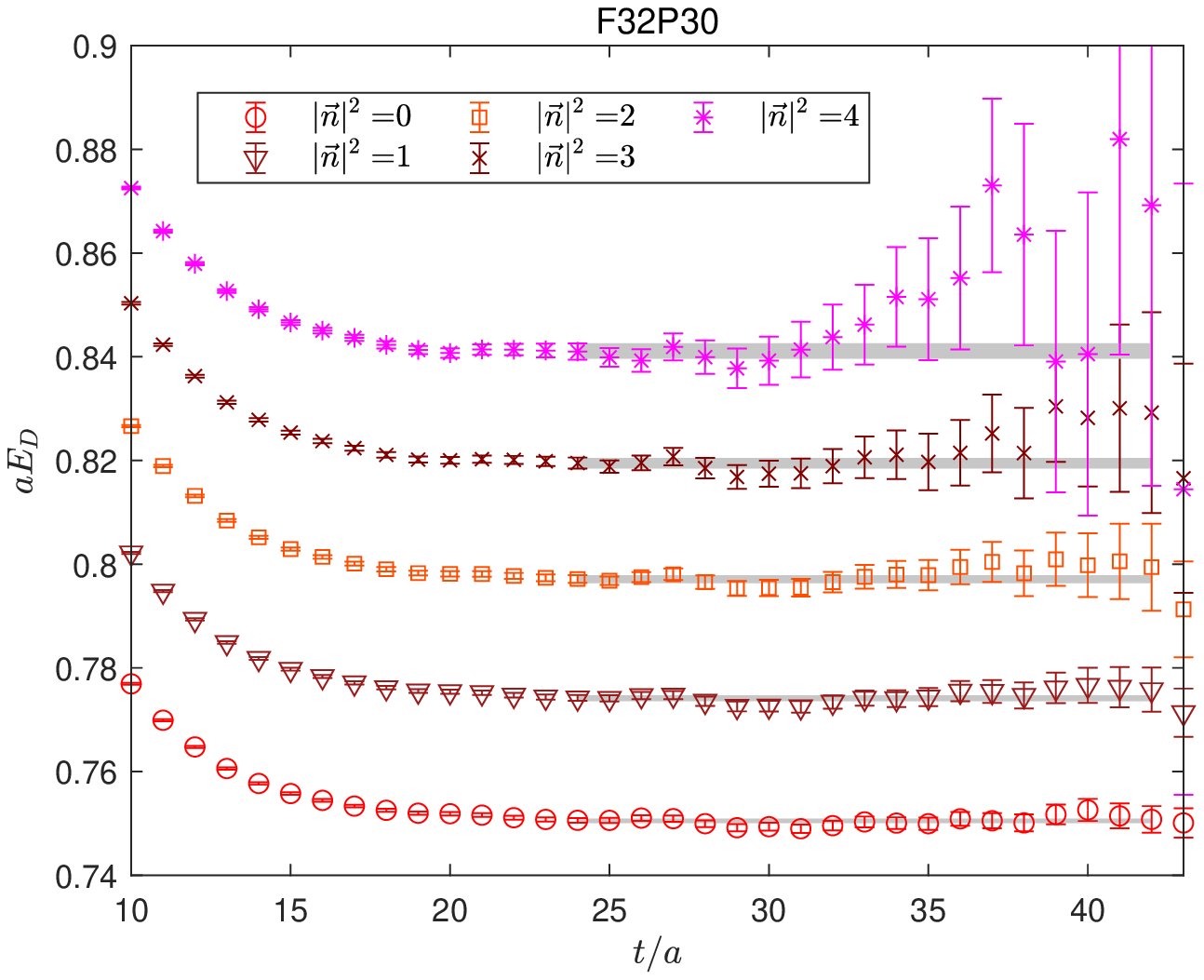}
}
\subfigure{
\centering
\includegraphics[width=8.5cm]{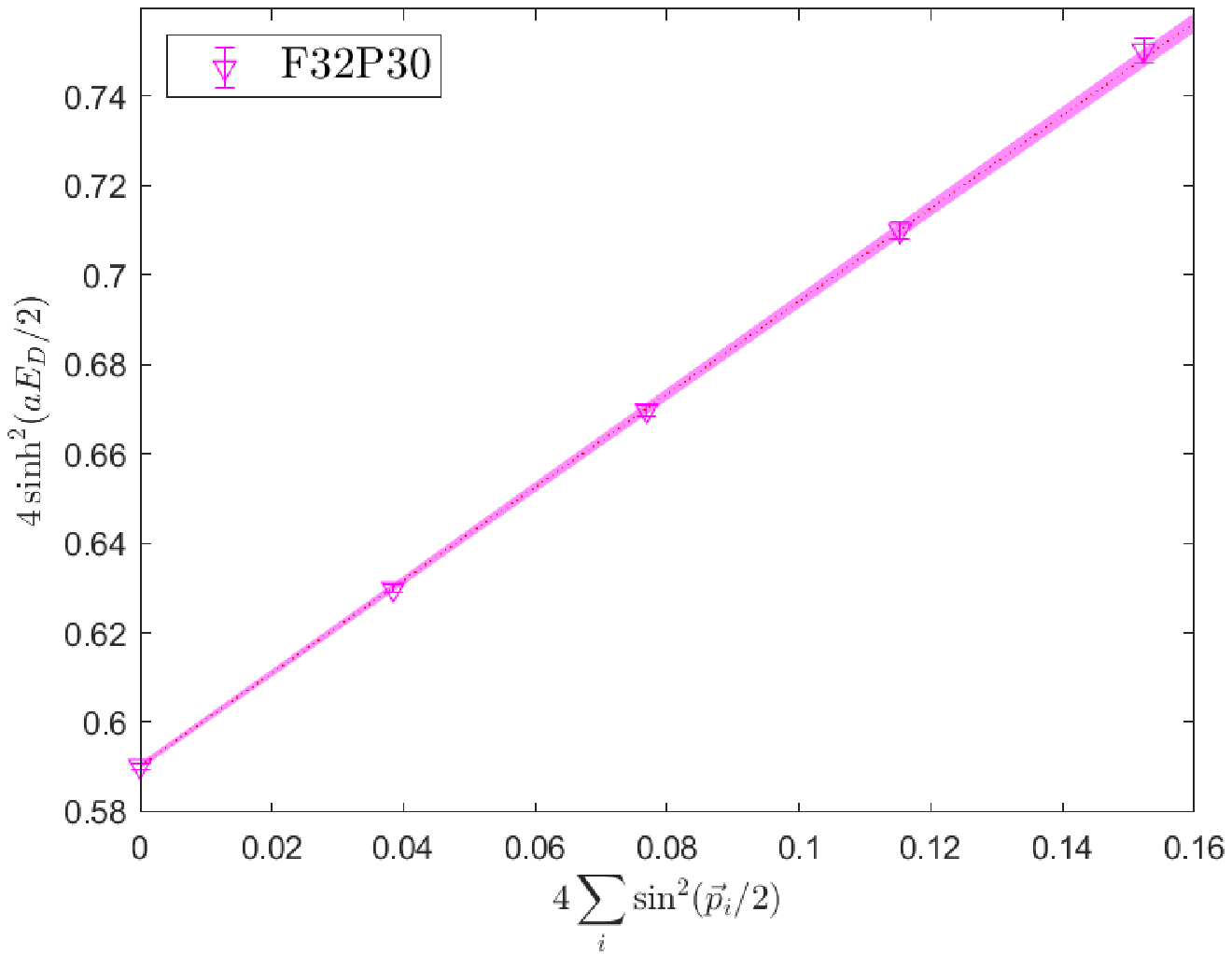}
}
\subfigure{
\centering
\includegraphics[width=8.5cm]{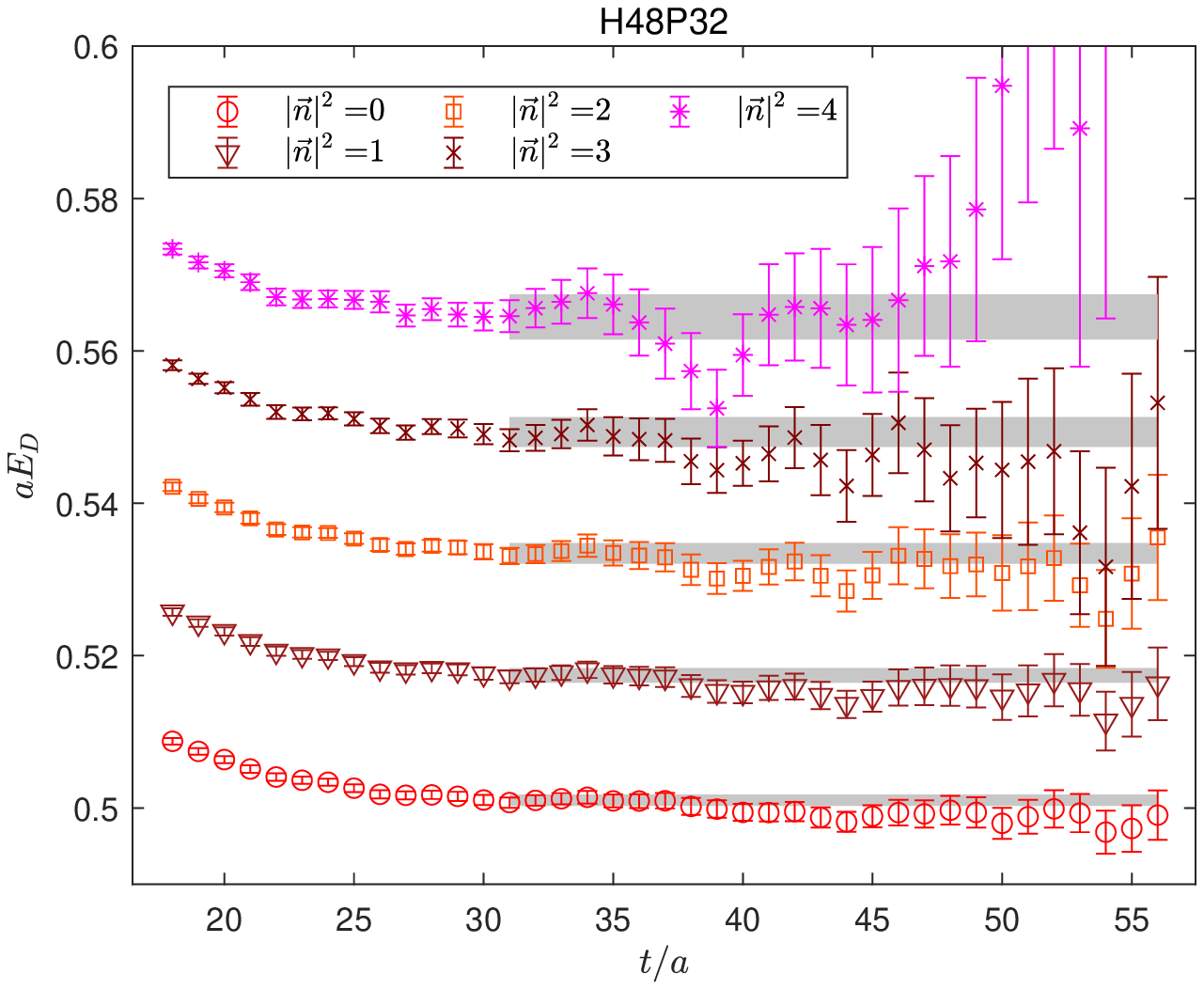}
}
\subfigure{
\centering
\includegraphics[width=8.5cm]{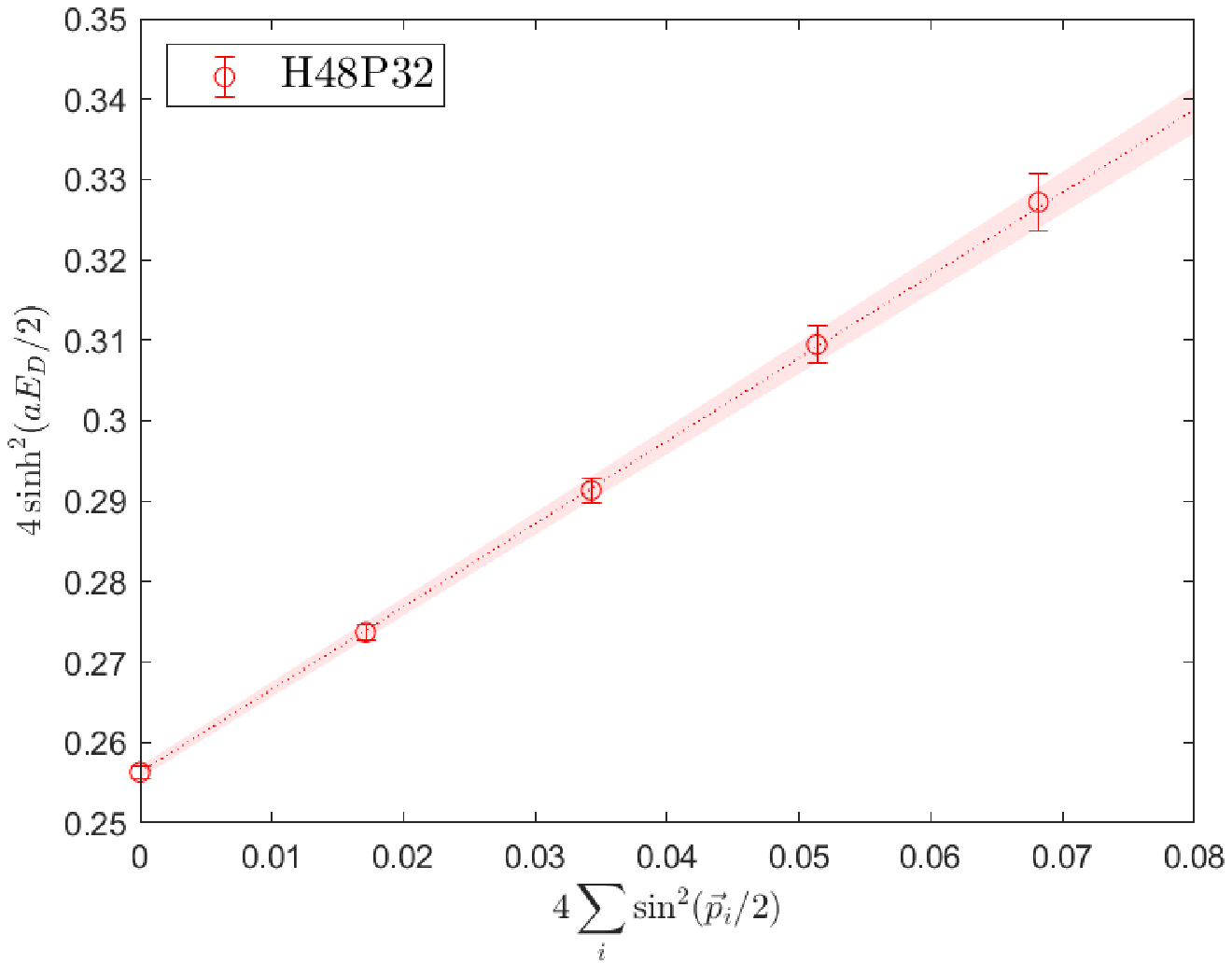}
}
\caption{\label{diag:mass} For the ensembles of C24P29, F32P30, and H48P32, the energy levels of $D$ particle with different momentum $\vec{p}=2\pi\vec{n}/L,|\vec{n}|^2=0,1,2,3,4$ extracted from two-point functions calculated using point source propagators(left) and the dispersion relation of $D$ meson(right). The horizontal gray bands in left panels denote the fitting ranges. 
The color regions in the right panels are the fitting results from Eq.~(\ref{eq:disper}). All the errors are purely statistical errors estimated by the jackknife method.}
\end{figure*}

\begin{table}[!h]
\begin{ruledtabular}
\begin{tabular}{cccc}
Ensemble & C24P29 & F32P30 &H48P32 \\
\hline
$M$(MeV) & 3098.6(3) & 3094.0(2) & 3096.6(6) \\
$Z_{M}$ & 0.2868(3) & 0.1718(2) & 0.0847(3) \\
\hline
$aE_{D}(|\vec{n}|^2=0)$ & 1.0257(8) & 0.7505(4) & 0.5010(7)\\
$aE_{D}(|\vec{n}|^2=1)$ & 1.0537(10) & 0.7747(5) & 0.5174(9) \\
$aE_{D}(|\vec{n}|^2=2)$ &1.0807(14) & 0.7971(7)& 0.5334(14) \\
$aE_{D}(|\vec{n}|^2=3)$ & 1.1068(19) & 0.8195(10) & 0.5493(20) \\
$aE_{D}(|\vec{n}|^2=4)$ & 1.1290(27) & 0.8411(15) & 0.5645(30) \\
$\mathcal{Z}_{\textrm{latt}}^{D}$ & 1.012(18) & 1.040(12) & 1.030(33) \\
$Z_{D}$ & 0.1028(8) & 0.0604(12) & 0.0274(9) \\
\hline
$aE_{D_s}(|\vec{n}|^2=0)$ & 1.0658(3) & 0.7801(2) & 0.5211(3)\\
$aE_{D_s}(|\vec{n}|^2=1)$ & 1.0928(4) & 0.8026(2) & 0.5369(5) \\
$aE_{D_s}(|\vec{n}|^2=2)$ &1.1191(5) & 0.8244(3)& 0.5525(8) \\
$aE_{D_s}(|\vec{n}|^2=3)$ & 1.1446(6) & 0.8456(5) & 0.5680(12) \\
$aE_{D_s}(|\vec{n}|^2=4)$ & 1.1674(8) & 0.8652(8) & 0.5842(20) \\
$\mathcal{Z}_{\textrm{latt}}^{D_s}$ & 1.035(6) & 1.026(6) & 1.043(21) \\
$Z_{D_s}$ & 0.2165(5) & 0.1376(3) & 0.0721(6) \\
\end{tabular}
\end{ruledtabular}
\caption{Mass spectra of the $J/\psi$ particle, energy levels $E_{D}$, $E_{D_s}$ with $\vec{p}=2\pi\vec{n}/L,|\vec{n}|^2=0,1,2,3,4$, overlap function $Z_{M/D/D_s}$, and the coefficients $\mathcal{Z}_{\textrm{latt}}^{D/D_s}$ defined by the discrete dispersion relation.}
\label{tab:meson_mass}
\end{table}

The energy levels of the particle $D/D_s$ are extracted from the two-point functions which are calculated by the point source propagators. A single-state correlated fit with the formula Eq.~(\ref{eq:2pt}) is utilized and the numerical fitting results of the spectra are summarized in Table~\ref{tab:meson_mass}. The effective levels of the particle $D$ are shown in Fig.~\ref{diag:mass} for all the ensembles and the horizontal gray bands therein denote the fitting center values and statistical errors estimated by the jackknife method. The dispersion relation of $D$ particle is also checked by using its energy levels. It is found that the discrete
dispersion relation
\be\label{eq:disper}
4\sinh^2\frac{E_{h}}{2} =4\sinh^2\frac{m_{h}}{2}+\mathcal{Z}_{\textrm{latt}}^{h}\cdot 4\sum\limits_i \sin^2 \frac{\vec{p}_i}{2}
\ee
describes the energies and momenta well and there is a nice linear behavior between $4\sinh^2 (E_{h}/2)$ and $4\sum\limits_i \sin^2 (\vec{p}_i/2)$ as illustrated in Fig.~\ref{diag:mass}. For the results of $D_s$ particle, one can refer to our previous paper~\cite{Meng:2024gpd}.

\subsection{Form factor}\label{sec:form_factor_result}
\begin{table}[!h]
\center
\begin{ruledtabular}
\begin{tabular}{cccccc}
Ensemble & C24P29 & F32P30 &H48P32 \\
\hline
$F_0^{D_s}(|\vec{n}|^2=0)$ &1.932(10) & 1.164(6) & 1.246(13) \\
$F_0^{D_s}(|\vec{n}|^2=1)$ & 1.731(8) & 1.459(4) & 1.111(10) \\
$F_0^{D_s}(|\vec{n}|^2=2)$ & 1.529(7) & 1.315(4) & 0.978(9)\\
$F_0^{D_s}(|\vec{n}|^2=3)$ & 1.340(6) & 1.450(4) & 0.857(9) \\
$F_0^{D_s}(|\vec{n}|^2=4)$ & 1.151(6) & 0.986(4) & 0.738(8) \\
$d_0^{D_s,(0)}$ & 1.229(7)  & 1.084(4)  & 0.812(9) \\
$d_0^{D_s,(1)}$ & 2.042(26) & 3.077(33) & 4.793(159) \\
$d_0^{D_s,(2)}$ &-0.021(81) & 0.303(169)& 3.16(2.02)\\
$\chi^2/\textrm{d.o.f}$ & 0.2 &0.6 & 0.04\\
\hline
$F_1^{D_s}(|\vec{n}|^2=0)$ & 1.350(5) & 1.161(3) & 0.870(8) \\
$F_1^{D_s}(|\vec{n}|^2=1)$ & 1.200(5) & 1.024(3) & 0.771(6) \\
$F_1^{D_s}(|\vec{n}|^2=2)$ & 1.740(5) & 0.908(3) & 0.691(6)\\
$F_1^{D_s}(|\vec{n}|^2=3)$ & 0.973(5) & 0.817(3) & 0.628(6) \\
$F_1^{D_s}(|\vec{n}|^2=4)$ & 0.876(5) & 0.739(3) & 0.571(6) \\
$d_1^{D_s,(0)}$ & 0.914(5)  & 0.784(3)  & 0.604(6) \\
$d_1^{D_s,(1)}$ & 0.965(20) & 1.464(19) & 2.208(114) \\
$d_1^{D_s,(2)}$ & 0.857(67) & 2.813(108)&10.29(1.54)\\
$\chi^2/\textrm{d.o.f}$ & 1.2 &0.6 & 0.2 \\
\hline
$F_2^{D_s}(|\vec{n}|^2=0)$ & 0.183(12)& 0.125(7) & 0.048(14) \\
$F_2^{D_s}(|\vec{n}|^2=1)$ & 0.102(9) & 0.048(6) & 0.020(10) \\
$F_2^{D_s}(|\vec{n}|^2=2)$ & 0.056(9) & 0.005(5) & 0.005(9)\\
$F_2^{D_s}(|\vec{n}|^2=3)$ & 0.040(8) &-0.006(5) & 0.003(10) \\
$F_2^{D_s}(|\vec{n}|^2=4)$ & 0.047(7) & 0.006(5) & 0.010(10) \\
$d_2^{D_s,(0)}$ & 0.041(7)  & -0.005(5) & 0.004(10) \\
$d_2^{D_s,(1)}$ &-0.134(30) &-0.223(30) &-0.276(140) \\
$d_2^{D_s,(2)}$ & 1.574(101)& 4.838(168)& 9.12(1.85)\\
$\chi^2/\textrm{d.o.f}$ &0.08 & 0.1 & 0.01 \\
\hline
$F_3^{D_s}(|\vec{n}|^2=0)$ & 1.166(12) & 1.034(7)& 0.921(17) \\
$F_3^{D_s}(|\vec{n}|^2=1)$ & 1.099(9) & 0.977(5) & 0.787(12) \\
$F_3^{D_s}(|\vec{n}|^2=2)$ & 1.016(8) & 0.902(5) & 0.688(12)\\
$F_3^{D_s}(|\vec{n}|^2=3)$ & 0.929(7) & 0.818(4) & 0.624(11) \\
$F_3^{D_s}(|\vec{n}|^2=4)$ & 0.826(7) & 0.729(4) & 0.576(11) \\
$d_3^{D_s,(0)}$ & 0.870(7)  & 0.785(4) & 0.601(11) \\
$d_3^{D_s,(1)}$ & 1.109(28) & 1.723(36) &1.842(140) \\
$d_3^{D_s,(2)}$ &-0.730(91) &-2.107(205)& 21.78(2.40)\\
$\chi^2/\textrm{d.o.f}$ &0.1 & 0.3& 0.1\\
\end{tabular}
\end{ruledtabular}
\caption{For the ensembles of C24P29, F32P30, and H48P32, the numerical results of $F_i^{D_s}(q^2)$, the polynomial fitting coefficient $d_i^{D_s,(j)}$ for $\vec{p}=2\pi\vec{n}/L(|\vec{n}|^2=0,1,2,3,4), i=0,1,2,3$, and $j=0,1,2$.}
\label{tab:F_Ds_values}
\end{table}

\begin{table}[!h]
\center
\begin{ruledtabular}
\begin{tabular}{cccccc}
Ensemble & C24P29 & F32P30 &H48P32 \\
\hline
$F_0^{D}(|\vec{n}|^2=0)$ & 1.870(25) & 1.637(11)& 1.237(31) \\
$F_0^{D}(|\vec{n}|^2=1)$ & 1.668(17) & 1.436(8) & 1.105(21) \\
$F_0^{D}(|\vec{n}|^2=2)$ & 1.456(14) & 1.241(6) & 0.956(18)\\
$F_0^{D}(|\vec{n}|^2=3)$ & 1.259(13) & 1.067(6) & 0.800(17) \\
$F_0^{D}(|\vec{n}|^2=4)$ & 1.069(13) & 0.914(6) & 0.665(17) \\
$d_0^{D,(0)}$ & 1.066(17) & 0.953(8)   & 0.699(22) \\
$d_0^{D,(1)}$ & 2.070(85) & 2.806(68)  & 5.961(524) \\
$d_0^{D,(2)}$ &-0.069(231) & 1.750(349)&-4.39(5.83)\\
$\chi^2/\textrm{d.o.f}$ &0.1 & 0.5 & 0.2\\
\hline
$F_1^{D}(|\vec{n}|^2=0)$ & 1.234(10) & 1.069(5)& 0.820(11) \\
$F_1^{D}(|\vec{n}|^2=1)$ & 1.084(8) & 0.931(4) & 0.713(8) \\
$F_1^{D}(|\vec{n}|^2=2)$ & 0.960(8) & 0.818(4) & 0.629(8)\\
$F_1^{D}(|\vec{n}|^2=3)$ & 0.859(8) & 0.730(4) & 0.568(8) \\
$F_1^{D}(|\vec{n}|^2=4)$ & 0.764(9) & 0.660(4) & 0.515(8) \\
$d_1^{D,(0)}$ & 0.766(10)  & 0.677(5)  & 0.527(9) \\
$d_1^{D,(1)}$ & 0.869(61)  & 1.164(47) & 1.780(227) \\
$d_1^{D,(2)}$ & 0.797(160) & 3.048(231)& 12.18(2.64)\\
$\chi^2/\textrm{d.o.f}$ & 0.3&0.08 &0.2 \\
\hline
$F_2^{D}(|\vec{n}|^2=0)$ & 0.032(29)  & -0.007(14)& -0.108(48) \\
$F_2^{D}(|\vec{n}|^2=1)$ & -0.047(18) & -0.074(9) & -0.097(28) \\
$F_2^{D}(|\vec{n}|^2=2)$ & -0.084(15) & -0.101(8) & -0.068(20)\\
$F_2^{D}(|\vec{n}|^2=3)$ & -0.089(14) & -0.097(7) & -0.028(19) \\
$F_2^{D}(|\vec{n}|^2=4)$ & -0.066(14) & -0.073(7) & 0.0043(20) \\
$d_2^{D,(0)}$ &-0.0647(14) &-0.821(7)  &-0.004(20) \\
$d_2^{D,(1)}$ &-0.433(100) &-0.635(75) &-1.695(517) \\
$d_2^{D,(2)}$ & 1.719(315) & 4.530(441)& 6.22(6.73)\\
$\chi^2/\textrm{d.o.f}$ & 0.02& 0.2& 0.07\\
\hline
$F_3^{D}(|\vec{n}|^2=0)$ & 1.195(29) & 1.075(14)& 0.818(25) \\
$F_3^{D}(|\vec{n}|^2=1)$ & 1.123(18) & 1.005(9) & 0.672(14) \\
$F_3^{D}(|\vec{n}|^2=2)$ & 1.029(15) & 0.914(7) & 0.578(11)\\
$F_3^{D}(|\vec{n}|^2=3)$ & 0.931(14) & 0.817(7) & 0.518(10) \\
$F_3^{D}(|\vec{n}|^2=4)$ & 0.814(15) & 0.720(7) & 0.469(9) \\
$d_3^{D,(0)}$ & 0.813(17)  & 0.746(8)  & 0.480(10) \\
$d_3^{D,(1)}$ & 1.311(90)  & 1.896(76) & 1.229(299) \\
$d_3^{D,(2)}$ &-0.868(277) &-1.677(420)& 22.37(4.57)\\
$\chi^2/\textrm{d.o.f}$ & 0.05& 0.4& 0.7\\
\end{tabular}
\end{ruledtabular}
\caption{For the ensembles of C24P29, F32P30, and H48P32, the numerical results of $F_i^{D}(q^2)$ and the polynomial fitting coefficient $d_i^{D,(j)}$ for $\vec{p}=2\pi\vec{n}/L(|\vec{n}|^2=0,1,2,3,4), i=0,1,2,3$, and $j=0,1,2$.}
\label{tab:F_D_values}
\end{table}

\begin{figure*}[htbp]
\centering
\subfigure{
\centering
\includegraphics[width=8.5cm]{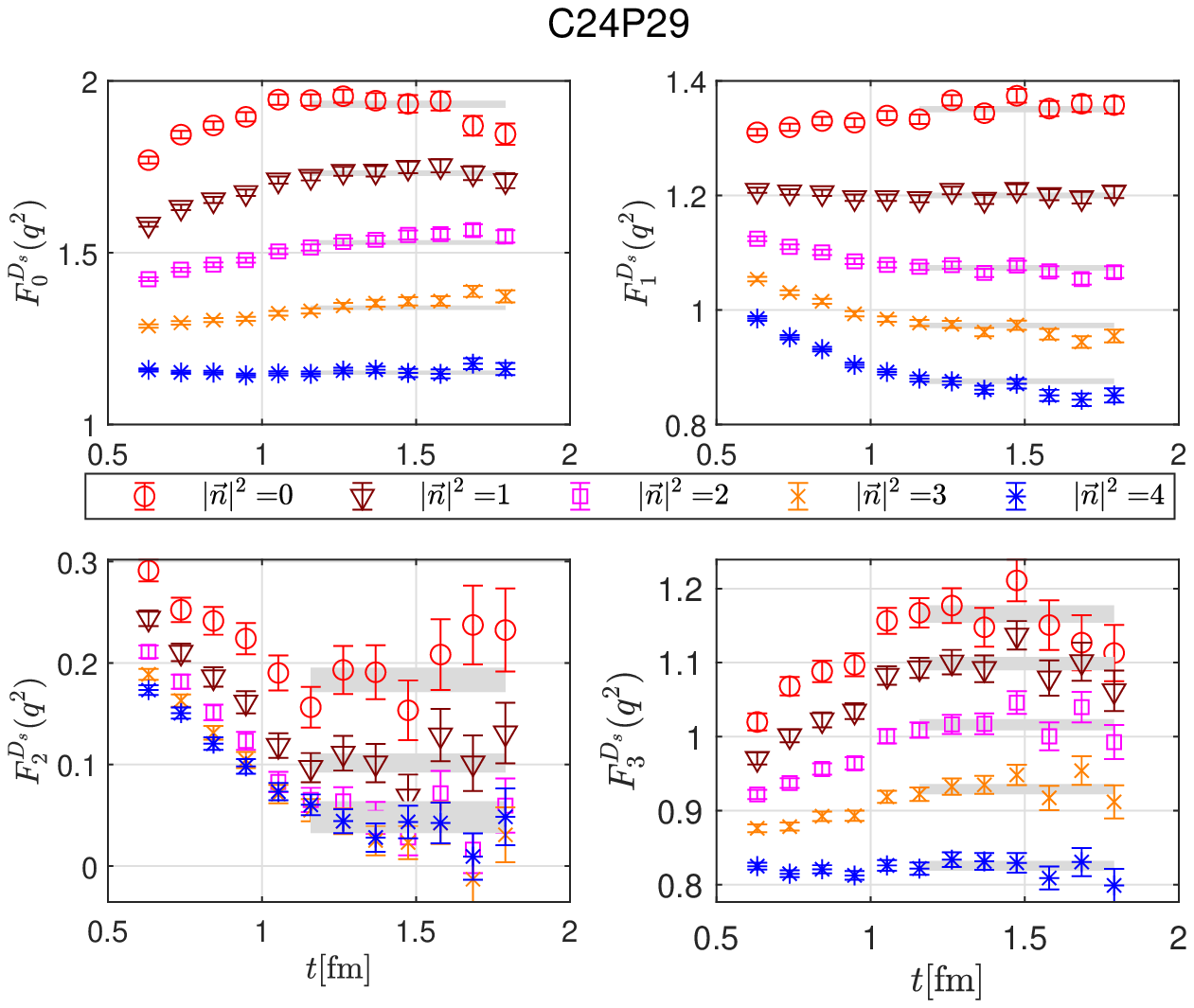}
}
\subfigure{
\centering
\includegraphics[width=8.5cm]{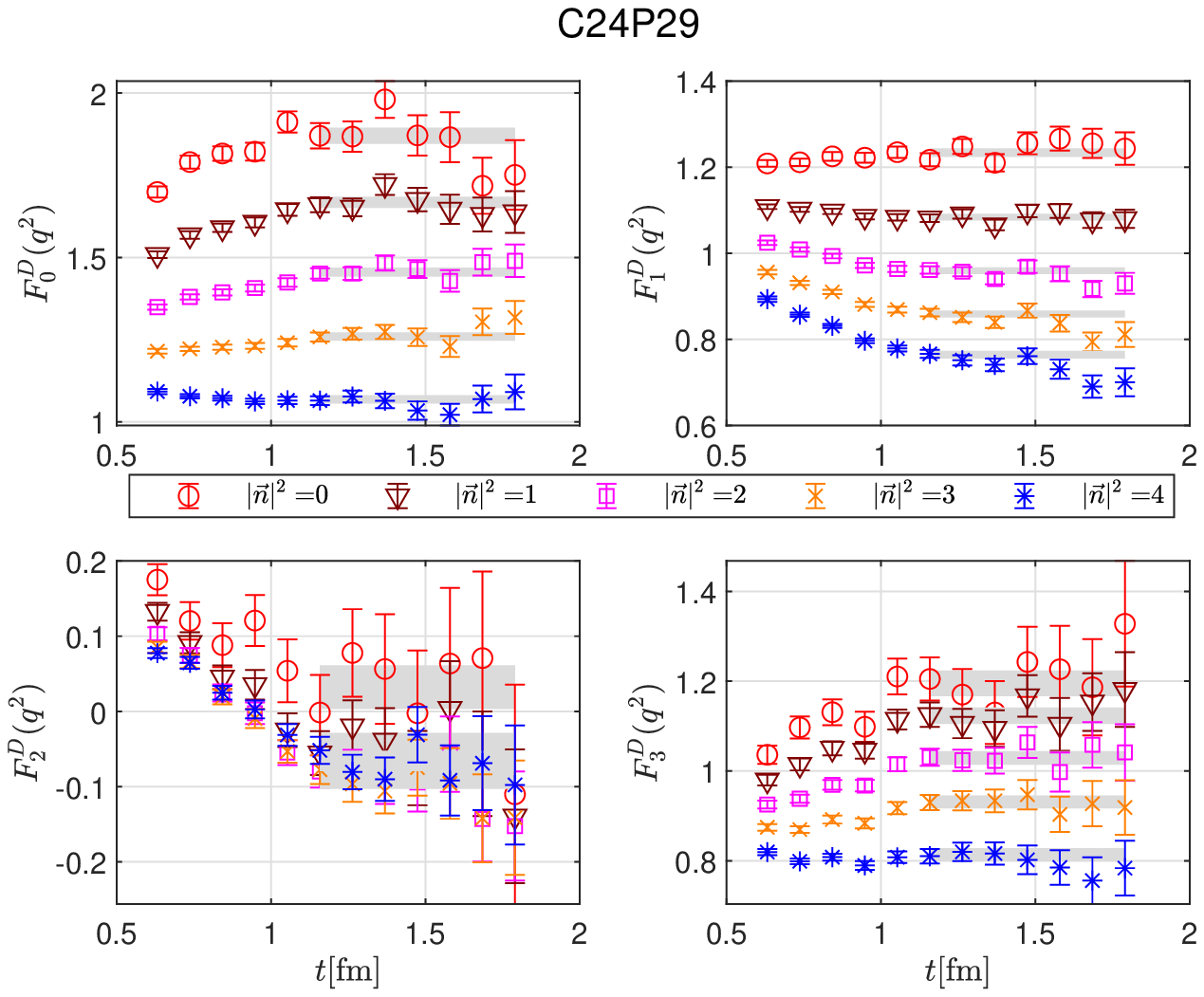}
}
\subfigure{
\centering
\includegraphics[width=8.5cm]{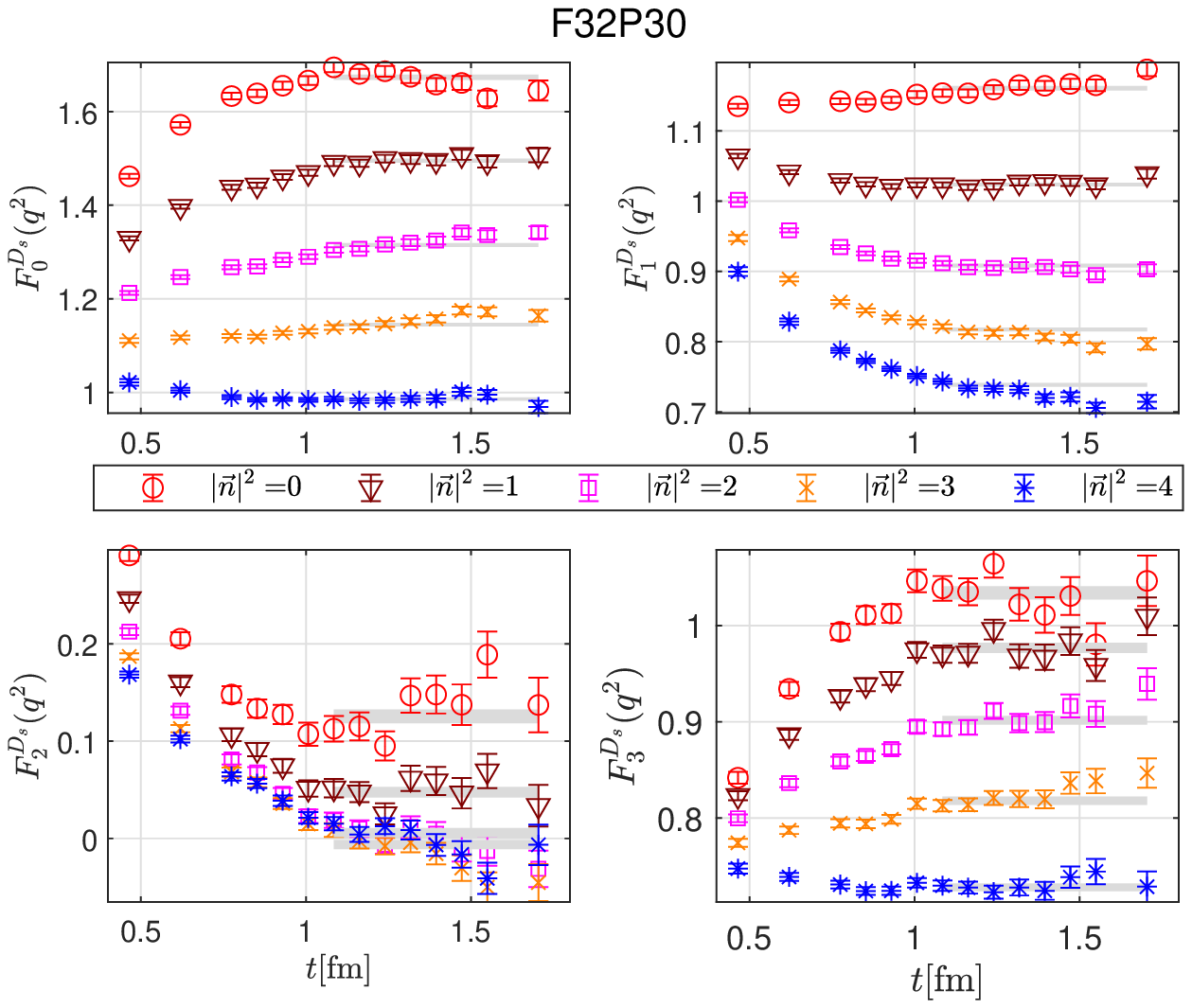}
}
\subfigure{
\centering
\includegraphics[width=8.5cm]{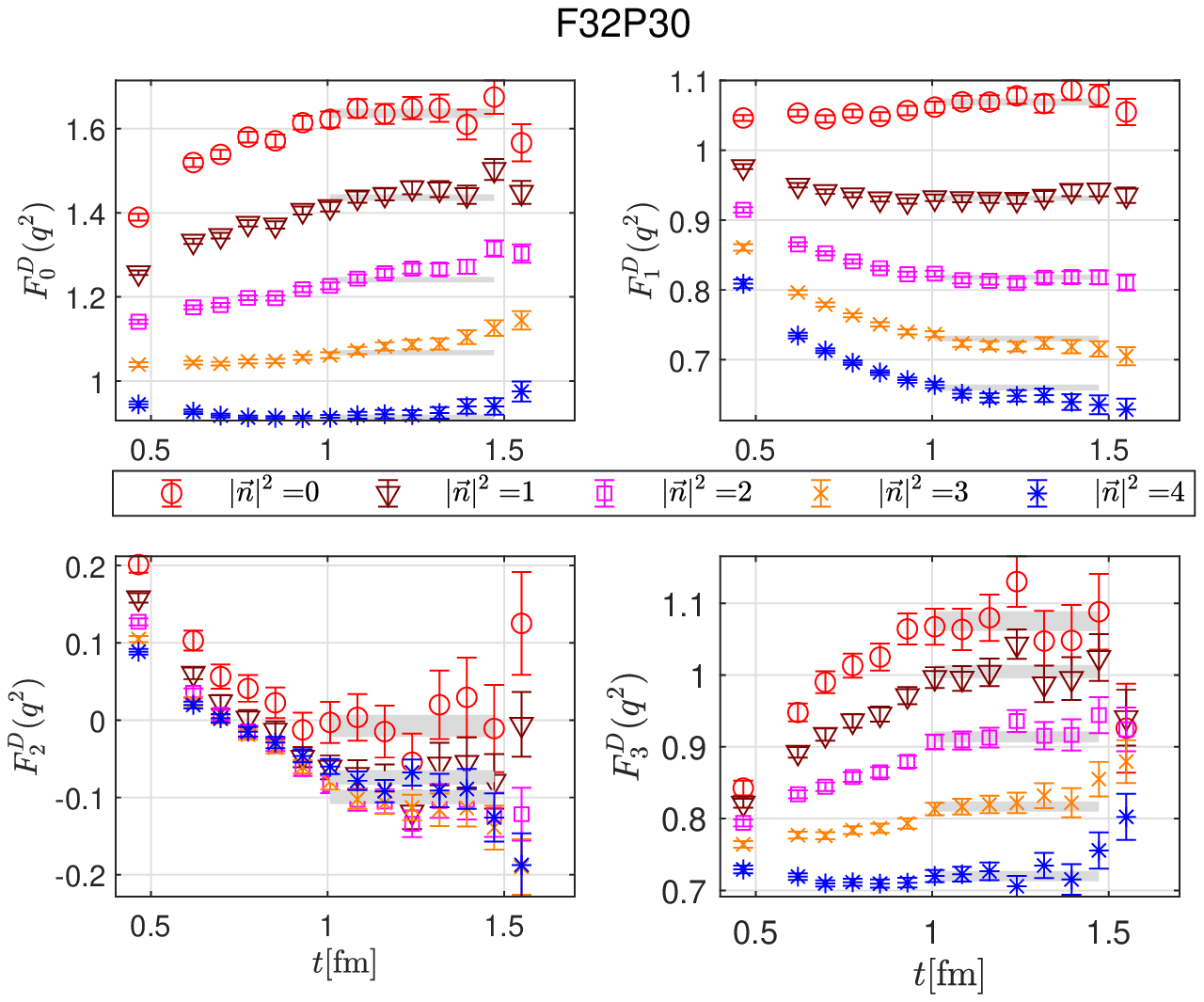}
}
\subfigure{
\centering
\includegraphics[width=8.5cm]{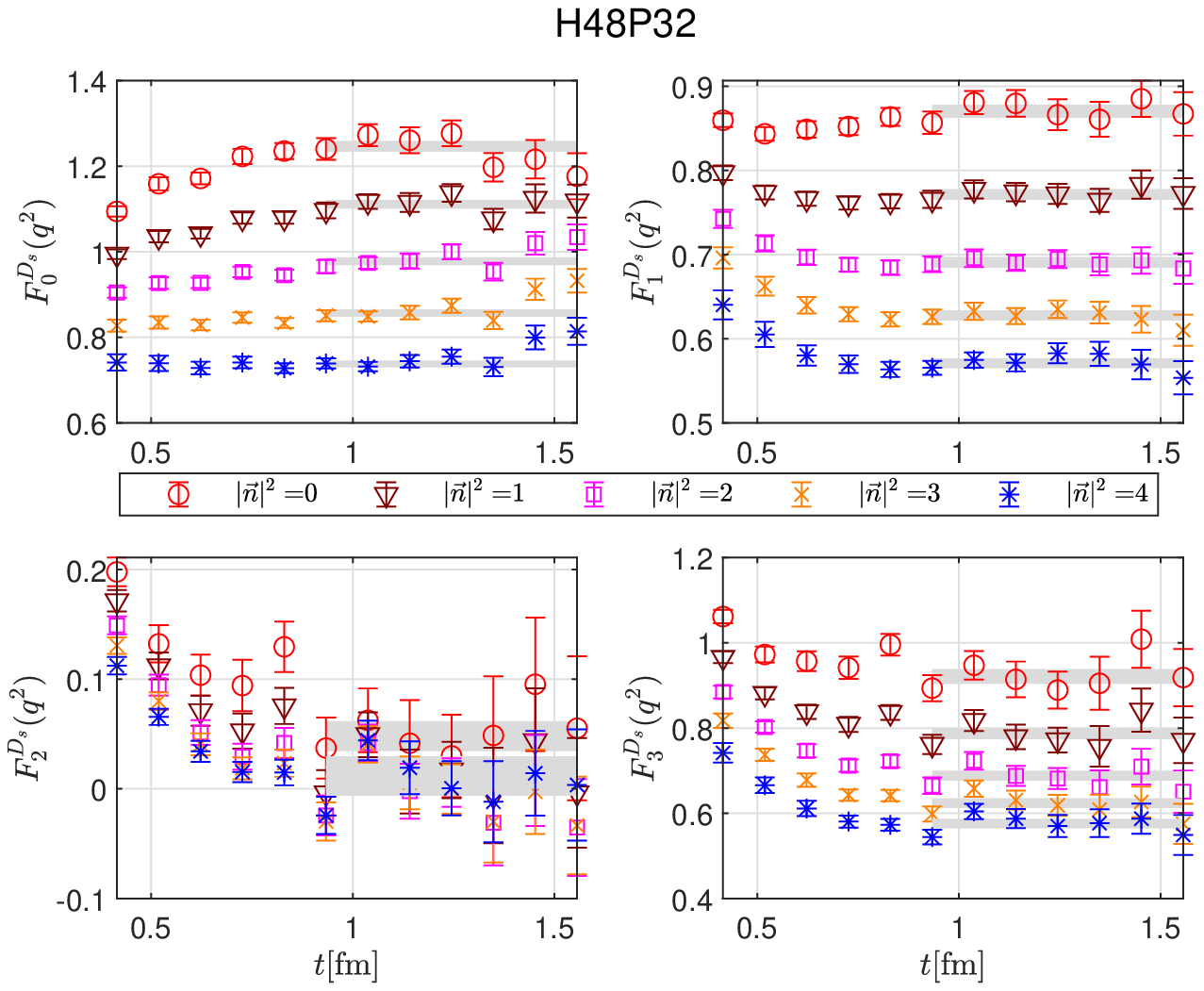}
}
\subfigure{
\centering
\includegraphics[width=8.5cm]{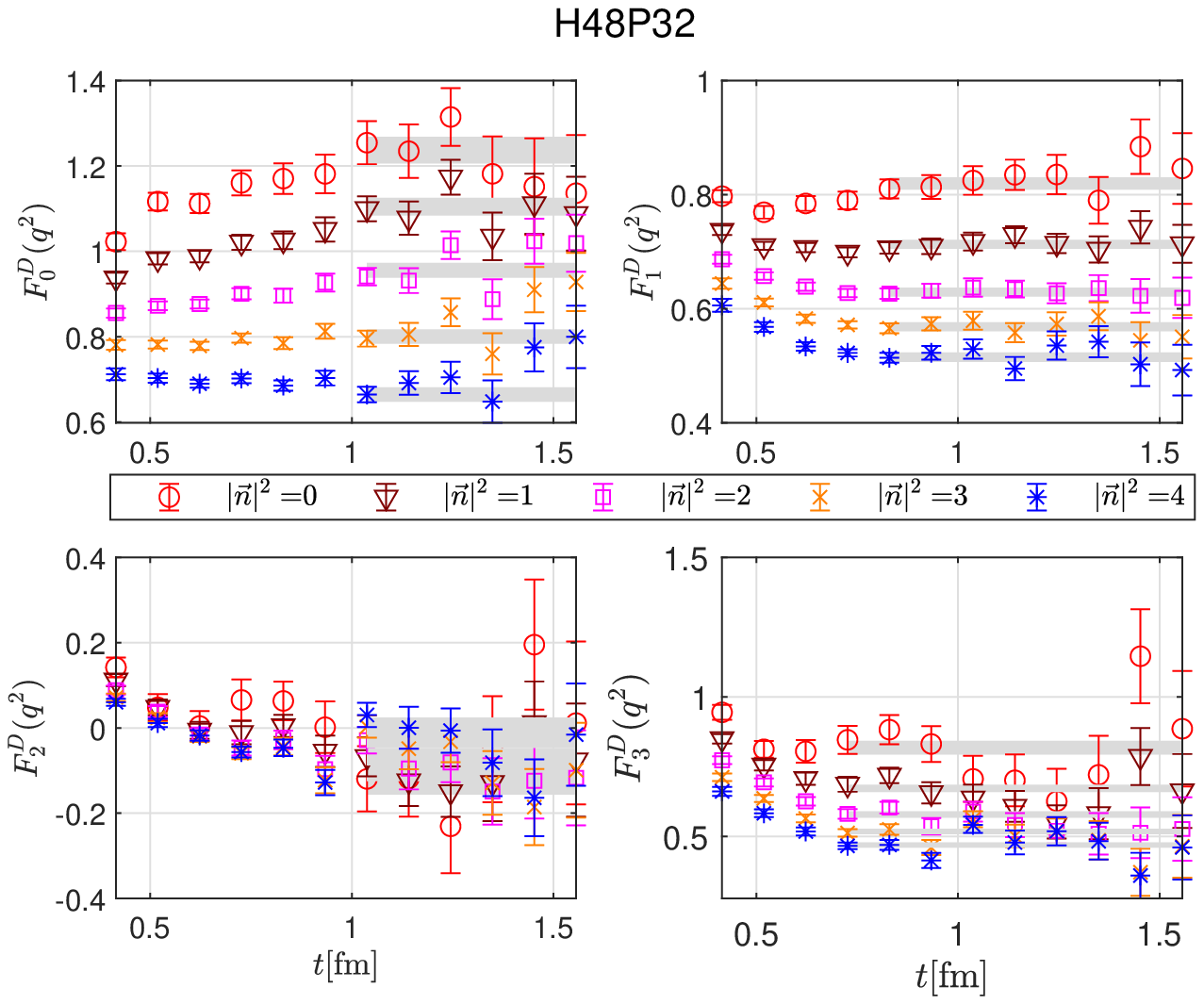}
}
\caption{\label{diag:F} For the ensembles of C24P29, F32P30, and H48P32, the form factors $F_i^{D_s}(q^2)$(left) and $F_i^{D}(q^2)$(right) with $\vec{p}=2\pi\vec{n}/L(|\vec{n}|^2=0,1,2,3,4)$ and $i=0,1,2,3$, where $F_i^{D_s}(q^2)$ denotes the form factor for the process $J/\psi\rightarrow D_sl\nu_l$, and $F_i^{D}(q^2)$ for $J/\psi\rightarrow Dl\nu_l$.}
\end{figure*}

\begin{figure*}[htbp]
\centering
\subfigure{
\centering
\includegraphics[width=16cm]{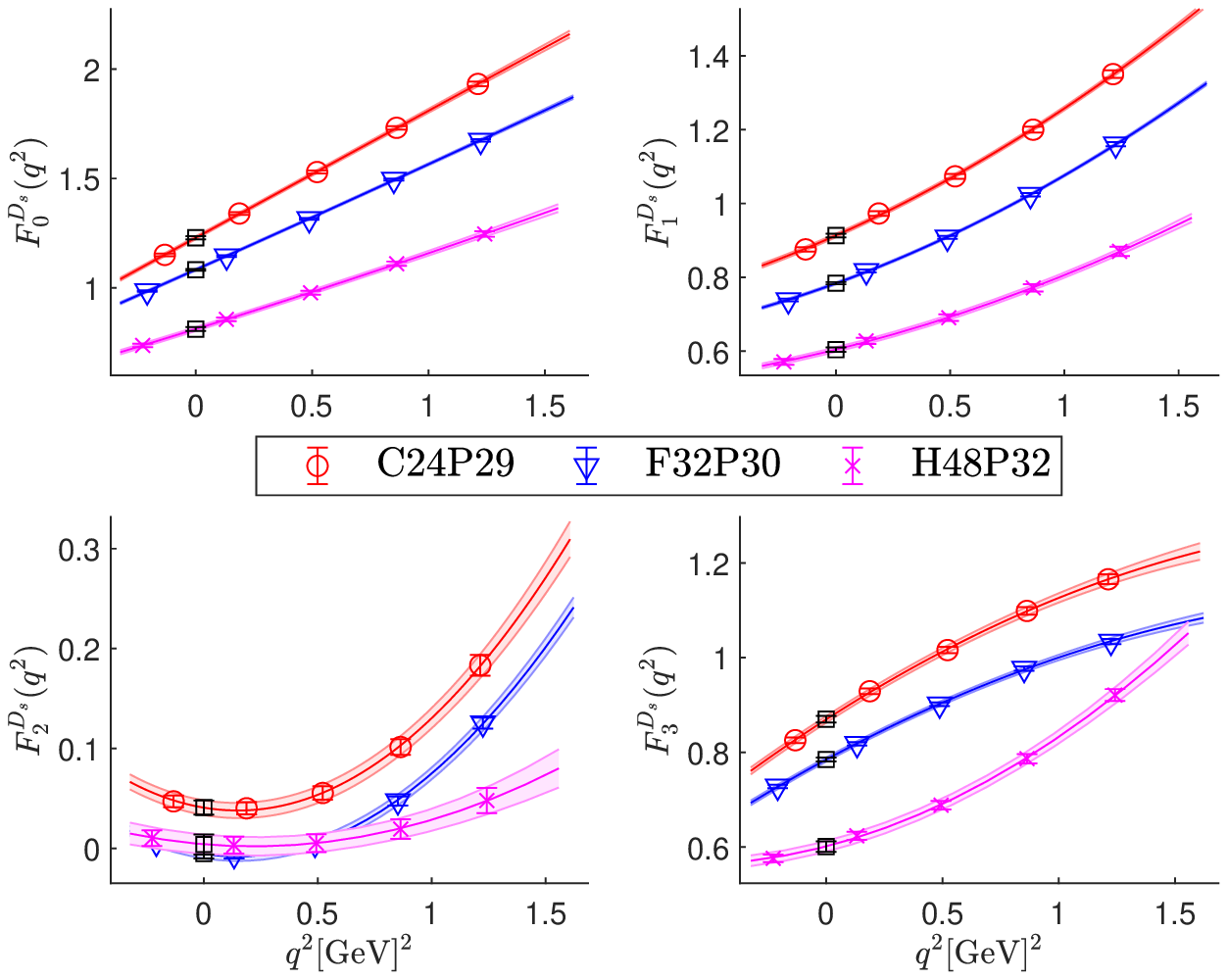}
}
\caption{\label{diag:F_cont} For the ensembles of C24P29, F32P30, and H48P32, the polynomial fits for the form factors $F_i^{D_s}(q^2)$ with $\vec{p}=2\pi\vec{n}/L(|\vec{n}|^2=0,1,2,3,4)$ and $i=0,1,2,3$, where $F_i^{D_s}(q^2)$ denotes the form factor for the process $J/\psi\rightarrow D_sl\nu_l$. The black boxes indicate the results at $q^2=0$.}
\end{figure*}

\begin{figure*}[htbp]
\centering
\subfigure{
\centering
\includegraphics[width=16cm]{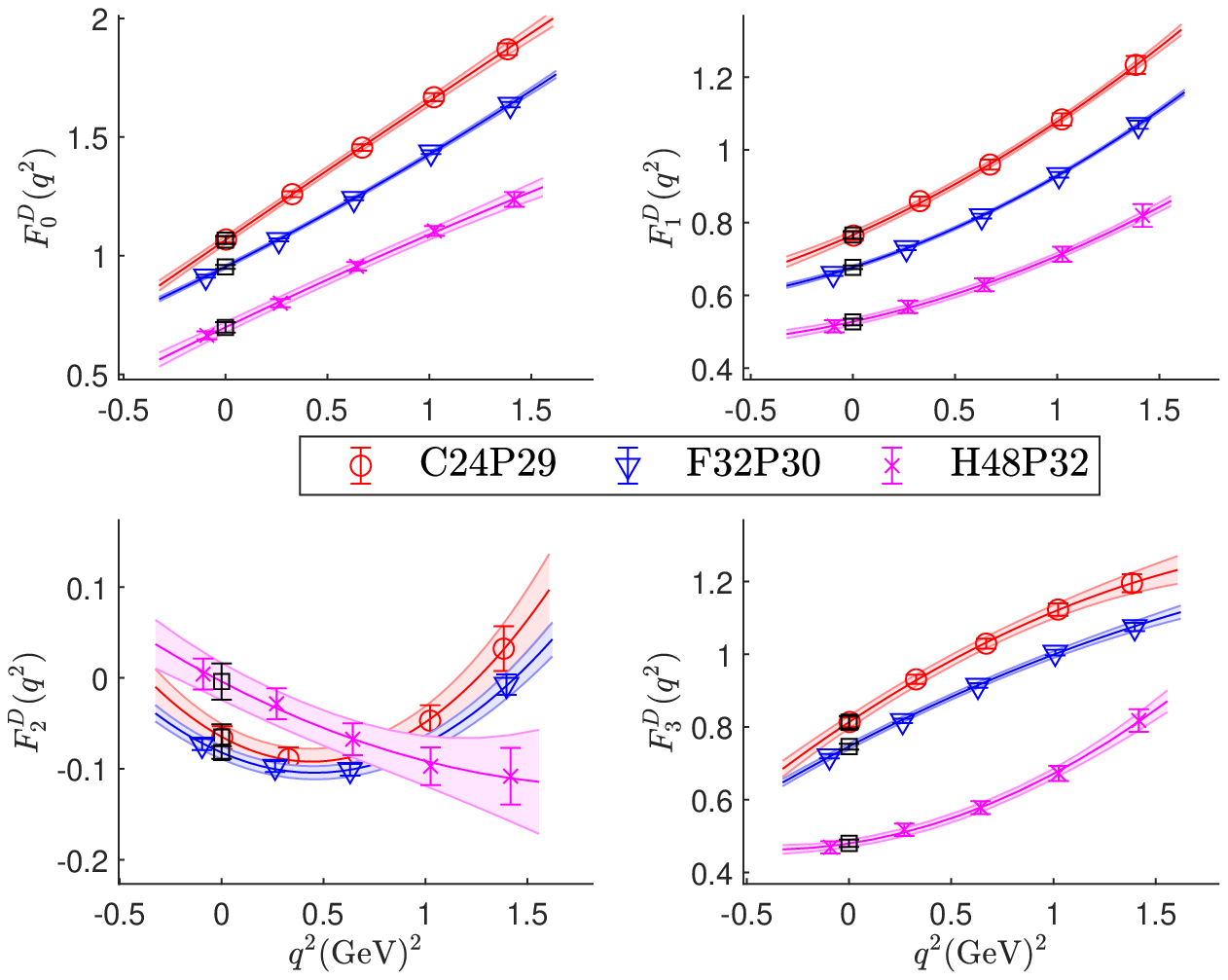}
}
\caption{\label{diag:F_cont_D} For the ensembles of C24P29, F32P30, and H48P32, the polynomial fits for the form factors $F_i^{D}(q^2)$(right) with $\vec{p}=2\pi\vec{n}/L(|\vec{n}|^2=0,1,2,3,4)$ and $i=0,1,2,3$, where $F_i^{D}(q^2)$ for $J/\psi\rightarrow Dl\nu_l$. The black boxes indicate the results at $q^2=0$.}
\end{figure*}

\begin{figure*}[htbp]
\centering
\subfigure{
\centering
\includegraphics[width=8.5cm]{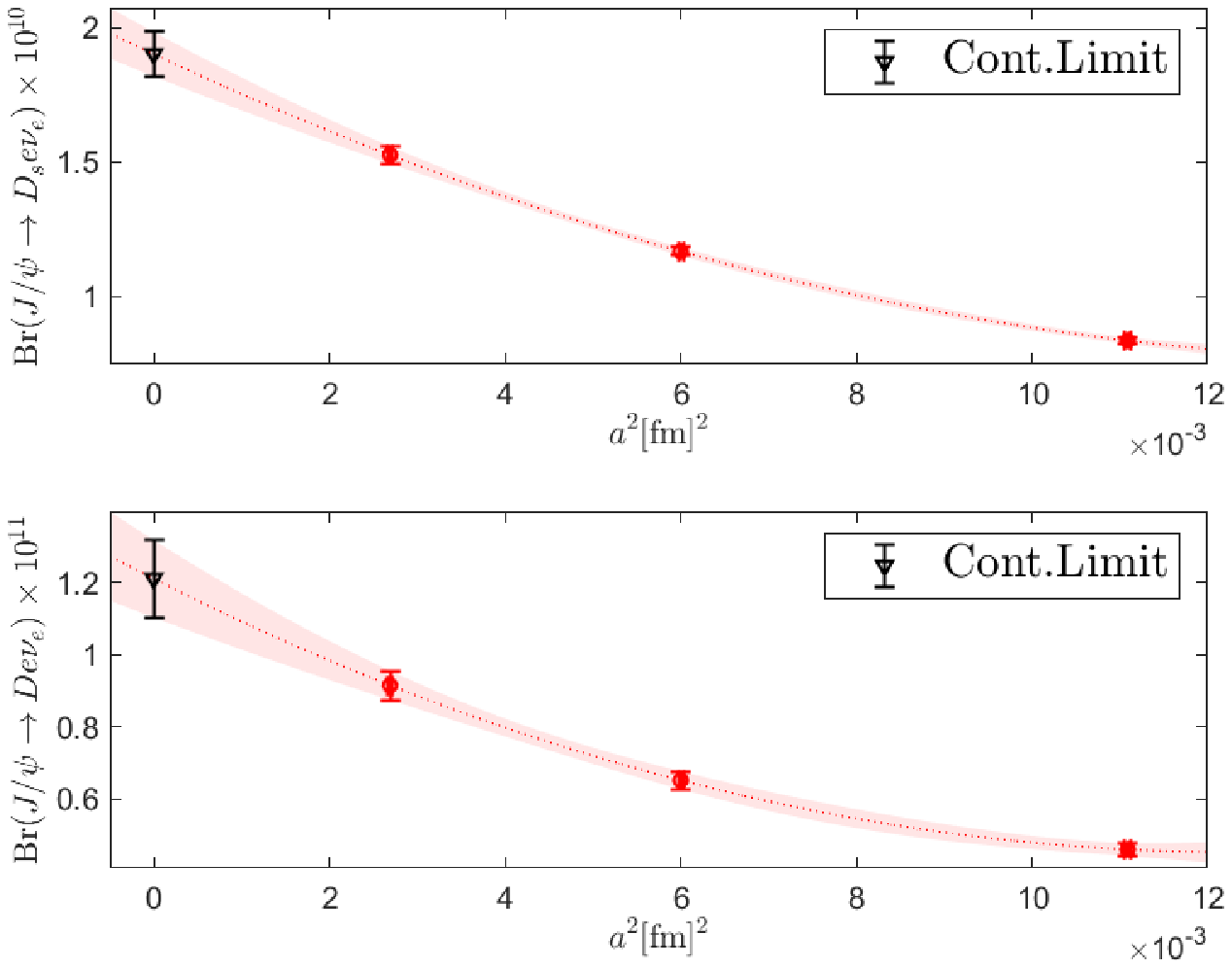}
}
\subfigure{
\centering
\includegraphics[width=8.5cm]{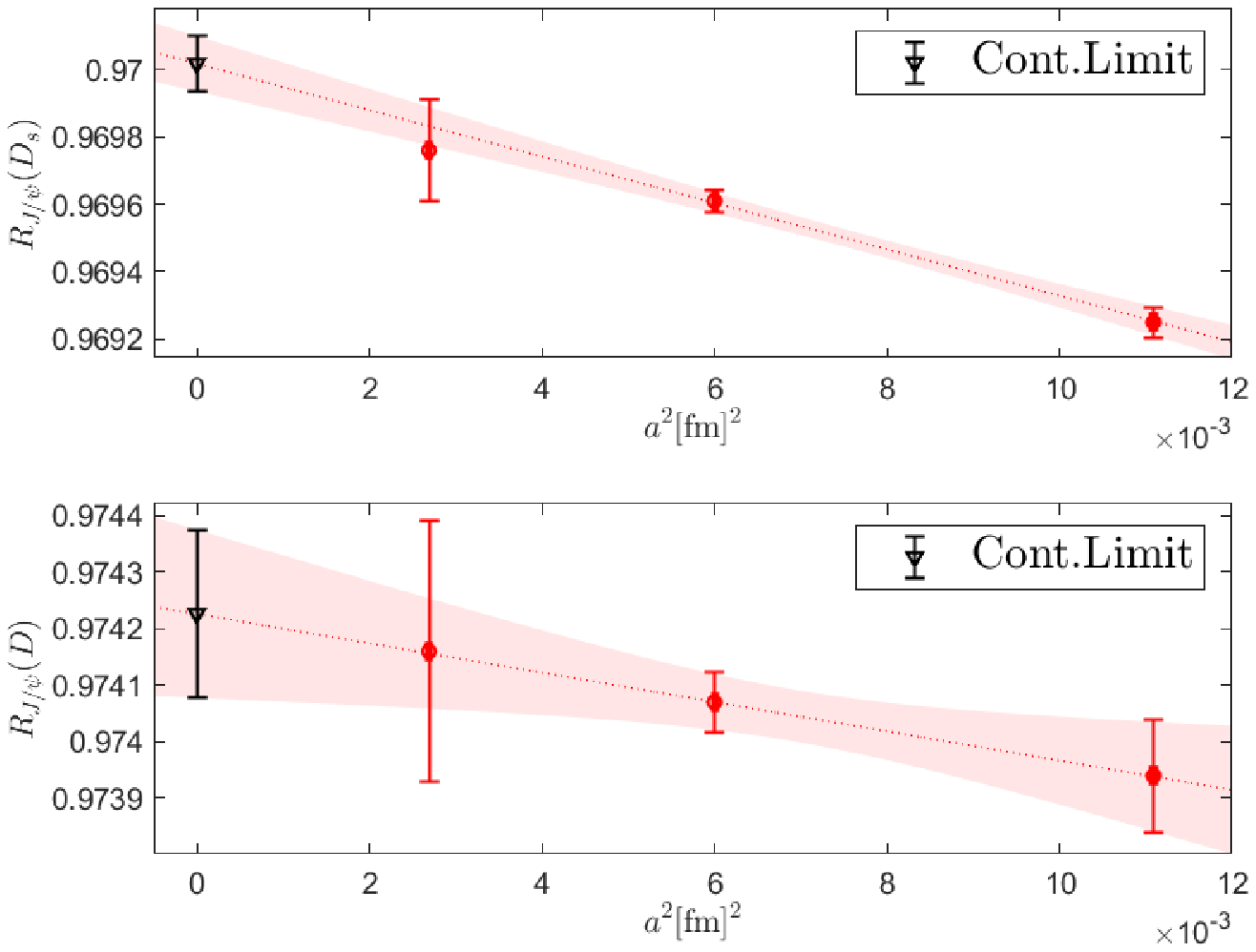}
}
\caption{\label{diag:Gamma_cont} The lattice results of $\operatorname{Br}(J/\psi\rightarrow D/D_s e\nu_e)$(left) and $R_{J/\psi}(D/D_s)$(right) as a function of the lattice spacings. On the left, a continuum extrapolation including $a^4$ term is performed. In the right, only a linear $a^2$ behavior is considered since the discretization effects are greatly canceled for the ratios $R_{J/\psi}(D)$ and $R_{J/\psi}(D_s)$ . The errors of lattice spacing have been included in the continuum limit, which are presented by the horizontal error bars.}
\end{figure*}
 The lattice results of $F_i^{D/D_s}(q^2)(i=0,1,2,3)$ as a function of the time separation $t$ are shown in Fig.~\ref{diag:F}, together with a series of momenta $\vec{p}=2\pi\vec{n}/L,|\vec{n}|^2=0,1,2,3,4$.
Here we present all the results of the three ensembles, i.e. C24P29, F32P30, and H48P32 from left to right, respectively. It shows that both $F_i^{D}$ and $F_i^{D_s}$ have obvious $t$ dependence as $t$ increases, indicating sizable excited-state effects associated with the initial and final states. With enough time intervals utilized in this work, we could observe obvious plateaus in a large enough time region. It is therefore natural to perform a correlated fit for these lattice data to a constant at a suitable time region. All results of $F_i^{D/D_s}(q^2)$ with different momenta $\vec{p}$ are obtained in this way and are denoted by the gray bands in the figure. 

With the five momentum modes of the form factor $F_i^{D/D_s}(q^2)$ taken into account,
a momentum extrapolation  can be performed by a polynomial form
\be\label{eq:mom_extra}
F_i^{h}(q^2)=d_i^{h,(0)}+d_i^{h,(1)}\cdot q^2+d_i^{h,(2)}\cdot q^4 
\ee
where the coefficients $d_i^{h,(j)}$ with $j=0,1,2$ are introduced. A correlated fit is adopted for the form and the covariance matrices of these five momentum modes are summarized in the Appendix.~\ref{sec:corr_mat}. As shown in Fig.~\ref{diag:F_cont} and Fig.~\ref{diag:F_cont_D} , such a polynomial formula in Eq.~(\ref{eq:mom_extra}) can describe the lattice data very well. In the following, we can see that the $q^2$ with five momentum modes $\vec{p}=2\pi\vec{n}/L,|\vec{n}|^2=0,1,2,3,4$ have covered the full phase space for all gauge ensembles in our calculation. The numerical results of the coefficients $d_i^{D,(j)}$ and $d_i^{D_s,(j)}$ are summarized in Table~{\ref{tab:F_Ds_values}} and Table~{\ref{tab:F_D_values}}. 

\begin{table*}[t]
\center
\begin{ruledtabular}
\begin{tabular}{ccccc}
Ensemble & C24P29 & F32P30 &H48P32 & $\textrm{Cont.Limit}$ \\
\hline
$\operatorname{Br}(J/\psi\rightarrow D_se\nu_e)\times 10^{10}$ & $0.837(7)_{\textrm{stat}}(10)_{V_{cs}}$ & $1.168(6)_{\textrm{stat}}(14)_{V_{cs}}$ & $1.526(26)_{\textrm{stat}}(19)_{V_{cs}}$ &  $1.90(6)_{\textrm{stat}}(5)_{V_{cs}}$  \\
$\operatorname{Br}(J/\psi\rightarrow D_s\mu\nu_{\mu})\times 10^{10}$ & $0.813(7)_{\textrm{stat}}(10)_{V_{cs}}$ & $1.133(6)_{\textrm{stat}}(14)_{V_{cs}}$ & $1.480(25)_{\textrm{stat}}(18)_{V_{cs}}$ & $1.84(6)_{\textrm{stat}}(5)_{V_{cs}}$  \\
$\operatorname{Br}(J/\psi\rightarrow De\nu_e)\times 10^{11}$ & $0.459(7)_{\textrm{stat}}(17)_{V_{cd}}$ & $0.651(6)_{\textrm{stat}}(24)_{V_{cd}}$ & $0.915(24)_{\textrm{stat}}(33)_{V_{cd}}$ & $1.21(6)_{\textrm{stat}}(9)_{V_{cd}}$ \\
$\operatorname{Br}(J/\psi\rightarrow D\mu\nu_{\mu})\times 10^{11}$ & $0.447(7)_{\textrm{stat}}(16)_{V_{cd}}$ & $0.634(6)_{\textrm{stat}}(23)_{V_{cd}}$ & $0.892(23)_{\textrm{stat}}(32)_{V_{cd}}$ & $1.18(6)_{\textrm{stat}}(9)_{V_{cd}}$ \\
\end{tabular}
\end{ruledtabular}
\caption{Numerical results of the branching fraction $\operatorname{Br}(J/\psi\rightarrow  D/D_s l\nu_l)$ with $l=e,\mu$ for all the gauge ensembles, together with the physical results after the continuum limit. The first errors are the statistical errors from the lattice simulation, and the second ones come from the uncertainties of the CKM matrix $V_{cs(d)}$.}
\label{tab:width}
\end{table*}

\subsection{Decay width}\label{sec:decay_width_result}

Using the numerical values of the coefficients $d_i^{D/D_s,(j)}$ as inputs, and combining with the Eq.~(\ref{eq:width}), the differential decay rate $d\Gamma/(dq^2V_{cs(d)}^2)$ can be calculated directly with the continuous $q^2$ covering the full phase space. After integrating the $q^2$ in the full phase space and taking into account the values of $V_{cs}=0.975(6)$ and $V_{cd}=0.221(4)$~\cite{pdg2022}, we immediately obtain the total branching fraction of the corresponding decay channels $\operatorname{Br}(J/\psi\rightarrow D/D_s l\nu_l)$, numerical values of which are listed in the Table.~\ref{tab:width}.

The lattice results for the branching fraction $J/\psi \rightarrow D/D_s e\nu_e$ at different lattice spacings are shown in Fig.~\ref{diag:Gamma_cont}, together with the ratio of between lepton $\mu$ and $e$ denoted by $R_{J/\psi}(D/D_s)=\operatorname{Br}(J/\psi \rightarrow D/D_s \mu\nu_{\mu})/\operatorname{Br}(J/\psi \rightarrow D/D_s e\nu_e)$. It is observed the branching fraction can be described well by a continuous limit extrapolation that contains the $a^4$-order term, whereas the ratios only the $a^2$ term. This is because the ratio can greatly eliminate the additional discretization effects. The leading $a^2$ term is expected since the ensembles used in the work have adopted the tadpole-improved tree-level Symanzik gauge action and the tadpole-improved tree-level clover fermion action. The next-leading order $a^4$ term has a sizeable effect on the charmonium system especially for a coarse lattice spacing that is larger than 0.1 fm, for instance. The continuous extrapolatin including $a^4$ term has also been used in previous lattice studies on charmonium system~\cite{Davies:2010ip,Hatton:2020qhk}.

Since we only have three lattice spacing values, we cannot make the continuum
extrapolation that includes $a^4$ term in a controlled fashion. Instead, we can only ensure a well-controlled continuum limit in a linear $a^2$ behavior given three lattice values. Because of this, we will call the former a naive continuum extrapolation in this paper. Finally, we report the first lattice prediction as
\beq\label{eq:Br_value}
\operatorname{Br}(J/\psi \rightarrow D_s e\nu_e)&=&1.90(6)_{\textrm{stat}}(5)_{V_{cs}}\times 10^{-10} \nonumber \\
\operatorname{Br}(J/\psi \rightarrow D e\nu_e)&=&1.21(6)_{\textrm{stat}}(9)_{V_{cd}}\times 10^{-11} \nonumber \\
R_{J/\psi}(D_s)&=&0.97002(8)_{\textrm{stat}} \nonumber \\
R_{J/\psi}(D)&=&0.97423(15)_{\textrm{stat}} 
\eeq
where the first errors are the statistical errors, and the seconds result from the uncertainties of CKM matrix
element $V_{cs(d)}$. The errors of the branching fraction $\operatorname{Br}(J/\psi\rightarrow D/D_s l\nu_l)$ with the different lepton $l=\mu$ and $l=e$ almost completely cancel, eventually leading to an extremely high precision value.

Note that the light quark masses in our calculation are nonphysical, it may have an influence on the process $J/\psi\rightarrow Dl\nu_l$, where the light quarks appear as valence quarks. For the process $J/\psi\rightarrow D_sl\nu_l$ , the nonphysical light quark only occurs in the sea, so its effect is expected to be small~\cite{FlavourLatticeAveragingGroupFLAG:2021npn}. Other systematic errors that have not been seriously considered include the effects from the neglected disconnected diagrams and the quenching of the charm quark. All these effects could be studied in future systematic lattice studies using e.g. the gauge ensembles with physical pion mass, with charm sea quarks, and with more lattice spacings. 

\subsection{Differential decay width}\label{sec:diff_decay_width_result}
\begin{figure*}[t]
\centering
\subfigure{
\centering
\includegraphics[width=16cm]{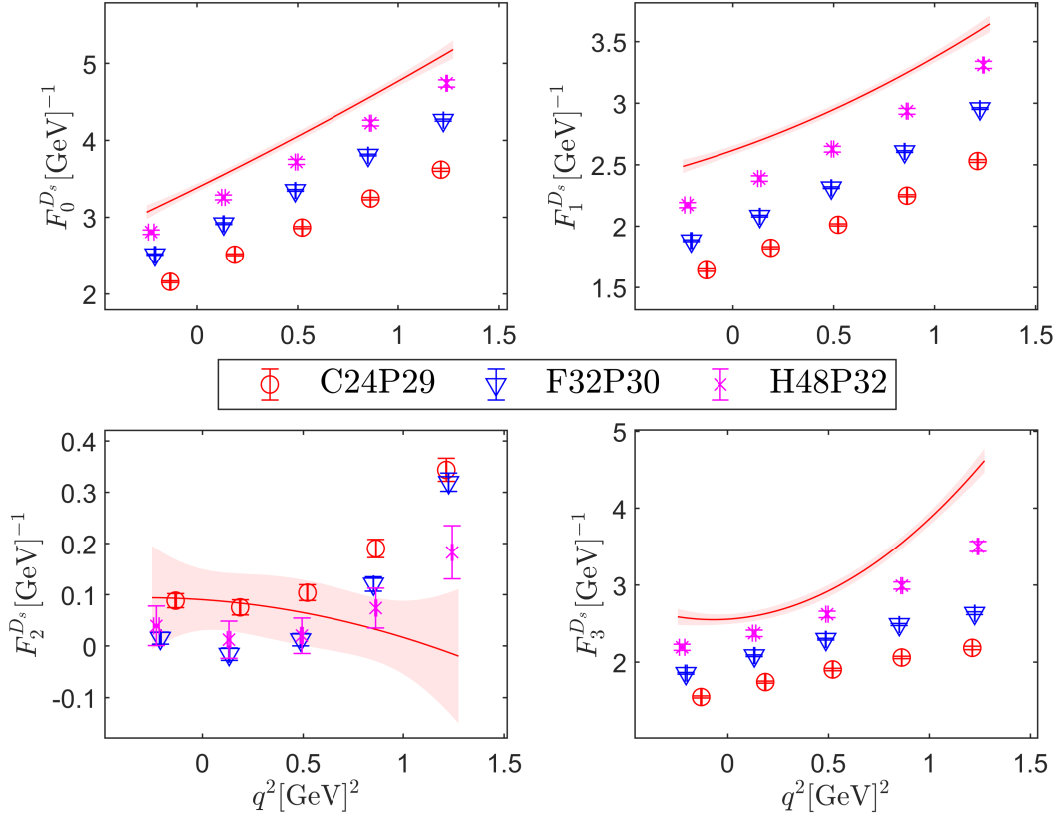}
}
\caption{\label{diag:F_cont_q} The lattice results of the $q^2$-expansion for $J/\psi\rightarrow D_s$ form factors. The red-shaded regions correspond to the final results in a continuum limit $a \rightarrow 0$.}
\end{figure*}

\begin{figure*}[t]
\centering
\subfigure{
\centering
\includegraphics[width=16cm]{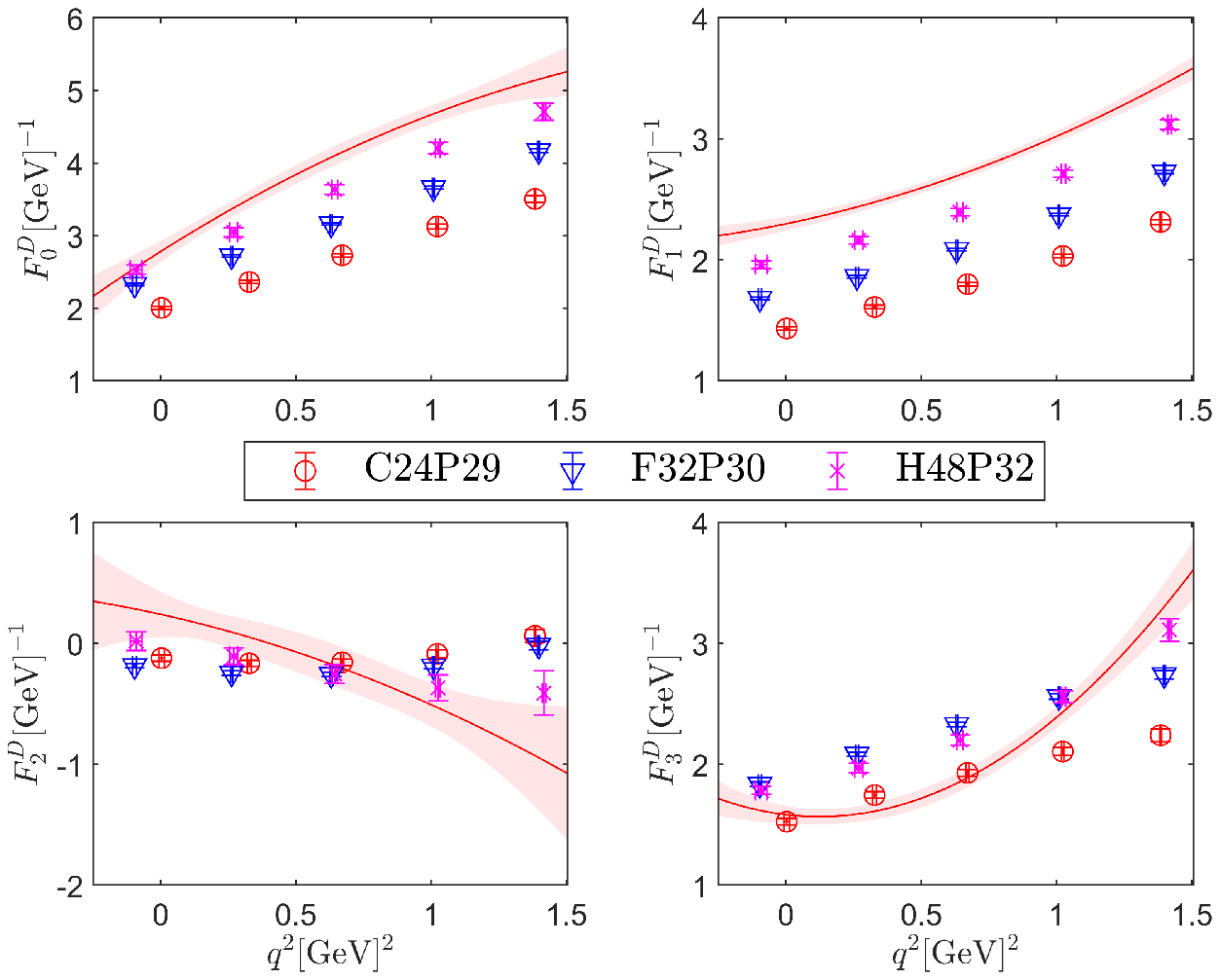}
}
\caption{\label{diag:F_cont_q_D} The lattice results of the $q^2$-expansion for $J/\psi\rightarrow D$ form factors. The red-shaded regions correspond to the final results in a continuum limit $a \rightarrow 0$.}
\end{figure*}

\begin{figure}[!h]
\centering
\includegraphics[width=8.5cm]{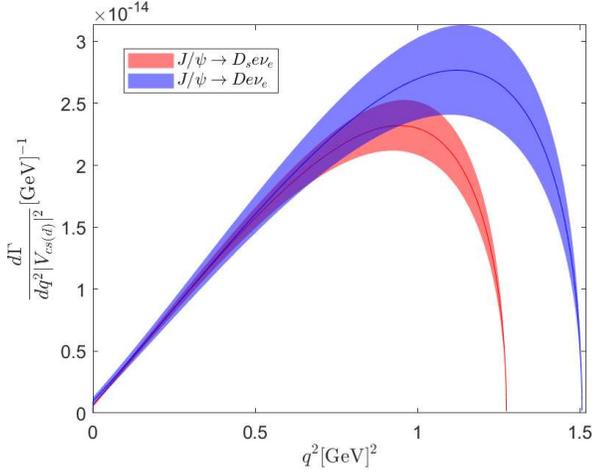}
\caption{\label{diag:width_cont_q} Predictions for the differential decay widths for $J/\psi\rightarrow D_s e \nu_e$ and $J/\psi\rightarrow D e \nu_e$, divided by $V_{cs}$ and $V_{cd}$, respectively.}
\end{figure}

\begin{table}[!h]
\center
\begin{ruledtabular}
\begin{tabular}{ccccc}
$J/\psi\rightarrow D_s/D$ & $c_0$ & $c_1$ & $c_2$ & $\chi^2/\textrm{d.o.f}$ \\
\hline
$F_0^{D_s}$ &3.38(5) &1.29(20) &0.10(19) & 1.2 \\
$F_1^{D_s}$ &2.62(4) & 0.57(11)&0.19(10)&  1.4\\
$F_2^{D_s}$ & 0.09(6)& -0.05(23)& -0.03(23)& 1.1\\
$F_3^{D_s}$ & 2.55(7)&  0.17(24) & 1.14(24)& 0.9\\
\hline
$F_0^D$ & 2.78(15)&2.37(58) &-0.48(46) & 0.9 \\
$F_1^D$ & 2.29(6)&0.46(13) & 0.27(9) & 1.2\\
$F_2^D$ & 0.24(18)& -0.51(83)& -0.25(69) & 1.3 \\
$F_3^D$ &1.58(7) &-0.25(30) & 1.06(26) & 0.9 \\
\end{tabular}
\end{ruledtabular}
\caption{Numerical results of the $q^2$-expansion coefficients $c_i,i=0,1,2$ for both processes $J/\psi\rightarrow D_s l \nu_l$ and $J/\psi\rightarrow D l \nu_l$.}
\label{tab:c_coef_cont}
\end{table}

To further compare with future experimental and phenomenological studies, in this section, we will give the results of the differential $q^2$ distribution of the decay width, i.e. $d\Gamma/dq^2$. For this purpose, we utilize a general $q^2$-expansion parametrization of form factors. In the above section, it has been found the polynomial $q^2$-expansion describes the lattice data very well, as shown in Fig.~\ref{diag:F_cont}. Therefore, the formula is given by
\be
F(a^2,q^2)=\sum\limits_{n=0}^{n_{\textrm{max}}}(c_n+d_na^2+f_na^4)q^{2n}
\ee
where the terms $d_n$ and $f_n$ describe the discretization effects. The $ a^4$ term is found to be necessary in the above discussions due to the coarse lattice spacing of the ensemble C24P29, thus we include it here. The $q^2$ distribution of these form factors are shown in Fig.~\ref{diag:F_cont_q} and Fig.~\ref{diag:F_cont_q_D}. The numerical values of coefficients $c_0,c_1$ and $c_2$ are collected in the Table~\ref{tab:c_coef_cont}. 

Taking into account the PDG values: $m_{J/\psi}=3.09690(1) \textrm{GeV}$, $m_{D_s}=1.96834(7) \textrm{GeV}$, and $m_{D}=1.86966(5) \textrm{GeV}$, one can obtain the differential decay width $d\Gamma/dq^2$, as shown in Fig.~\ref{diag:width_cont_q}. Integrating the $q^2$, it immediately leads to the branching fraction $\operatorname{Br}(J/\psi \rightarrow D_s e\nu_e)=1.99(14)\times 10^{-10}$
and $\operatorname{Br}(J/\psi \rightarrow D e\nu_e)=1.44(17)\times 10^{-11}$, where the errors are total errors including the uncertainties both from the statistics and CKM matrix element $V_{cs(d)}$. It is seen the results obtained here are almost compatible with Eq.~(\ref{eq:Br_value}), but with larger statistical errors. Therefore, we will take the values in Eq.~(\ref{eq:Br_value}) as our final reports.

\section{Discussion}\label{sec:discussion}
The hadronic matrix element of $J/\psi\rightarrow D/D_s$ should have the similar lorentz structure with other semileptonic decay $P\rightarrow V$, for example, $D_s\rightarrow \phi$~\cite{Donald:2013pea}, $B\rightarrow D^*$~\cite{Harrison:2017fmw}, and $B\rightarrow J/\psi$~\cite{Harrison:2020gvo}. Therefore, the method proposed in this work can be directly extended to the above processes. Previous lattice studies on these channels have utilized another parameterization scheme~\cite{Richman:1995wm}
\beq\label{eq:V_A_trad}
&&H_{\mu}(p',q)=\epsilon_{\mu\alpha\beta\delta}\epsilon_{\alpha}(p',\lambda)p'^{\beta}p^{\delta}\frac{2V(q^2)}{M+m} \nonumber \\
&+& 2MA_0(q^2)\frac{\epsilon(p',\lambda)\cdot q}{q^2} q_{\mu} \nonumber \\
&+&(M+m)A_1(q^2)\left[ \epsilon_{\mu}(p',\lambda)-\frac{\epsilon(p',\lambda)\cdot q}{q^2} q_{\mu}\right] \nonumber \\
&-&A_2(q^2)\frac{\epsilon(p',\lambda)\cdot q}{M+m}\left[p'_{\mu}+p_{\mu}-\frac{M^2-m^2}{q^2}q^{\mu}\right] \nonumber \\
\eeq
where form factors $V$ and $A_i$ are introduced, rather than $F_i$ in this work. The kinematic constraint gives 
\be\label{eq:kin_con}
A_0(0)=\frac{m+M}{2M}A_1(0)-\frac{M-m}{2M}A_2(0)
\ee
Note that the form factors with a factor of $q_{\mu}$ in Eq.~(\ref{eq:V_A_trad}) have no significant contribution to the experimental rate, due to its combination with the leptonic
current tends to zero as the lepton mass vanishes. The form factors $F_i$ introduced in this work can be related to the conventional form factors $V$ and $A_i$ by the following way

\beq
V&=&\frac{(m+M)}{2 m M}F_0 \nonumber \\
A_0&=&\frac{1}{2M}\left(F_1-\frac{M^2-m^2+q^2}{2 m M}F_2-\frac{M^2-m^2-q^2}{2m^2}F_3\right) \nonumber \\
A_1&=&\frac{F_1}{m+M}, \quad A_2= \frac{m+M}{2m^2M}(m F_2+M F_3) 
\eeq

In our simulations, we find $F_2(0)\simeq 0$, and it naturally leads to the well-known relationship given in Eq.~(\ref{eq:kin_con}). Given the $q^2$-expansion coefficients $c_i$, the results of $V(0)$ and $A_i(0)$ are determined directly, numerical values of which are summarized in Table.~\ref{tab:V_A_con}. 

\begin{table}[!h]
\center
\begin{ruledtabular}
\begin{tabular}{ccccc}
  & $V(0)$ & $A_0(0)$ & $A_1(0)$ & $A_2(0)$    \\
\hline
$J/\psi\rightarrow D_s$ & 1.40(2) & 0.11(1) & 0.52(1) & 1.71(5) \\
$J/\psi\rightarrow D$   & 1.19(6) &0.13(2)  & 0.46(1) &1.23(9)\\
\end{tabular}
\end{ruledtabular}
\caption{Numerical results of the conventional form factors $V(0),A_i(0)$.}
\label{tab:V_A_con}
\end{table}

The experimental search for $J/\psi$ semileptonic decay has been ongoing for many years, but it has not been detected to date. The highest upper limit comes from BESIII experiment on $J/\psi \rightarrow D e\nu_e $, where the full $1.01\times 10^{10}$ $J/\psi$ samples are utilized. However, there are still three orders of magnitude away from the theoretical prediction for the process $J/\psi \rightarrow D e\nu_e$. Since there exist no measurements using the full $J/\psi$ samples on $J/\psi \rightarrow D_s l\nu_l $ so far, we can deduce from previous measurements that the upper limit of the channel $J/\psi \rightarrow D_s e\nu_e $ might be $10^{-8}$ if all $J/\psi$ samples are used. The experimental results of $J/\psi$ semileptonic decay are summarized in Table.~\ref{tab:BESIII}, together with the lattice results of this work for comparisons. Nowadays, the future experiment, such as Super Tau Charm Facility~\cite{Achasov:2023gey}, has the great potential to improve the upper limit to a level of $10^{-10}$. By combining the certain experimental measurement of the branching fraction and the form factors given by lattice QCD, one can extract the CKM matrix element $V_{cs(d)}$, therefore providing an alternative precision test for the standard model.

\begin{table}[!h]
\center
\begin{ruledtabular}
\begin{tabular}{cccc}
 channels & Upper limit/Br & $J/\psi$ number      &  Refs        \\
\hline
$J/\psi\rightarrow D_s e\nu_e$ & $4.9\times 10^{-5}$ & $5.8\times 10^{7}$ & \cite{BES:2006mls}      \\
$J/\psi\rightarrow D_s e\nu_e$ & $1.3\times 10^{-6}$ & $2.3\times 10^{8}$  &  \cite{BESIII:2014pps}    \\
$J/\psi\rightarrow D e\nu_e$   & $7.1\times 10^{-8}$ &$1.01\times 10^{10}$ &  \cite{BESIII:2021mnd}    \\
$J/\psi\rightarrow D \mu\nu_{\mu}$ & $5.6\times 10^{-7}$  & $1.01\times 10^{10}$& \cite{BESIII:2023fqz}   \\
\hline
$J/\psi\rightarrow D_s e\nu_e$ & $1.90(8)\times 10^{-10}$ &  & this work    \\
$J/\psi\rightarrow D_s \mu\nu_{\mu}$ & $1.84(8)\times 10^{-10}$ &  & this work    \\
$J/\psi\rightarrow D e\nu_e$ & $1.21(11)\times 10^{-11}$ &   &  this work    \\
$J/\psi\rightarrow D \mu\nu_{\mu}$ & $1.18(11)\times 10^{-11}$ &   &  this work    \\
\end{tabular}
\end{ruledtabular}
\caption{ Upper limits of the branching fraction of $J/\psi \rightarrow D/D_s l\nu_l$ from BESIII experiments and the branching fractions(Br) predicted by lattice QCD.}
\label{tab:BESIII}
\end{table}

\section{Conclusion}\label{sec:conclude}
In this work, we present the first lattice QCD calculation on the semileptonic decay of $J/\psi$ particle. The weak decay $J/\psi\rightarrow D_sl\nu_l$ and $J/\psi \rightarrow D l \nu_l$ are studied respectively. The (2+1)-flavor Wilson-clover gauge ensembles with $m_{\pi }\sim 300$ MeV are utilized. After a native continuum limit under three lattice spacings, we finally obtain the lattice prediction $\operatorname{Br}(J/\psi\rightarrow D_s e\nu_e)=1.90(6)(5)_{V_{cs}}\times 10^{-10}$ and $\operatorname{Br}(J/\psi\rightarrow D e\nu_e)=1.21(6)(9)_{V_{cd}}\times 10^{-11}$ , where the first errors are statistical errors, and the seconds come from the uncertainties of CKM matrix element $V_{cs(d)}$. Given the form factors of $J/\psi\rightarrow D$ and $J/\psi\rightarrow D_s$, the ratios of the branching fractions between lepton $\mu$ and $e$ are computed immediately. The values of the ratios are given by $R_{J/\psi}(D_s)=0.97002(8)$ and $R_{J/\psi}(D)=0.97423(15)$ with only the statistical errors included. 
 
Combining with the exact experimental measurement in the future, the lepton flavor universality can be checked carefully and the CKM matrix element $V_{cs(d)}$ can also be extracted, thus providing an alternative precision test for the standard model. The lattice scheme adopted in this work can be directly applied to other $P\rightarrow V$ semileptonic decays, for example, $D\rightarrow K^*$~\cite{BESIII:2024jlj}, $B\rightarrow K^*$~\cite{LHCb:2023gpo}, and $B\rightarrow D^*$~\cite{HFLAV:2022esi}. 

\begin{acknowledgments}
We thank the CLQCD collaborations for providing us their gauge configurations with dynamical fermions~\cite{CLQCD:2023sdb}, which are generated on HPC Cluster of ITP-CAS, the Southern Nuclear Science Computing Center(SNSC), the Siyuan-1 cluster supported by the Center for High Performance Computing at Shanghai Jiao Tong University and the Dongjiang Yuan Intelligent Computing Center. The calculations were performed using the Chroma software suite~\cite{Edwards:2004sx} with QUDA~\cite{Clark:2009wm,Babich:2011np,Clark:2016rdz} through HIP programming model~\cite{Bi:2020wpt}. The authors are supported by NSFC of China under Grant No.12293060, No.12293063, No.12293062, No. 12305094, No.11935017, and the National Key R\&D Program of China No.2021YFB0300203. The numerical calculations are supported by the SongShan supercomputer at the National Supercomputing Center in Zhengzhou.
\end{acknowledgments}

\bibliography{ref}
\bibliographystyle{h-physrev}

\clearpage
\appendix
\section{Coefficients of $c_0,c_1$, and $c_2$ }\label{sec:c_coeff}
\begin{widetext}
\beq\label{eq:c_coef}
c_0&=& \frac{1}{m^4}\Big{[}-8E^3F_0^2m^2M+4E^3F_3m_l^2(2F_2m+F_3M)+4E^3M(F_2m+F_3M)^2 + 4E^2F_0^2m^2(m^2+3M^2+m_l^2)\nonumber \\
&+&2E^2m_l^2(m^2(F_2^2+F_3^2+4F_1F_3)- 2F_3F_2mM-2F_3^2M^2)-2E^2(F_2m+F_3M)(-4F_1m^2M+(F_2m+F_3M)(m^2+M^2)) \nonumber \\
&-&2E^2F_3^2m_l^4 - 4Em^2m_l^2(2F_0^2M+F_2m(F_1+2F_3)+F_3M(3F_1+F_3))-8EF_0^2m^2M^3-4EF_1^2m^4M \nonumber \\
&-&4EF_1F_2m^5-12EF_1F_2m^3M^2-4EF_1F_3m^4M-12EF_1F_3m^2M^3-4EF_2^2m^4M-8EF_2F_3m^3M^2\nonumber \\
&-&4EF_3^2m^2M^3+2F_0^2m^2((M^2+m_l^2)^2-m^4) + 2m^2(m^2+M^2)(F_1^2m^2+(2F_1M+F_2m+F_3M)(F_2m+F_3M))\nonumber \\
&-&2m^2m_l^2(m^2((F_1+F_3)^2+F_2^2)-2F_2mM(F_1+F_3)-2F_3M^2(2F_1+F_3))+2F_3m^2m_l^4(2F_1+F_3)\Big{]}
\nonumber \\
&+&\frac{8}{m}F_0F_1\left((E-M)(2ME-M^2-m^2-m_l^2)\right) \nonumber \\
c_1&=&\frac{1}{m^4}\Big{[} -8E^3(F_2m+F_3M)^2-8E^2\big{(}2F_0^2m^2M+2F_1m^2(F_2m+F_3M)-(F_2m+F_3M)(F_2mM+F_3(M^2+m_l^2))\big{)}   \nonumber \\ 
&+&8Em^2\big{(}F_0^2(m^2+3M^2+m_l^2)-F_1^2m^2+4F_1M(F_2m+F_3M)+F_1F_3m_l^2+(F_2m+F_3M)^2 \big{)} \nonumber \\
&+&8m^2M \big{(}-F_0^2(m^2+M^2)+F_1^2m^2-2F_1M(F_2m+F_3M)-(F_2m+F_3M)^2\big{)}\nonumber \\
&-& 8m^2m_l^2(M(F_0^2+F_3(2F_1+F_3))+F_2m(F_1+F_3)) \Big{]}  + \frac{16}{m}F_0F_1(2ME-M^2-m^2)\nonumber \\
c_2&=&\frac{1}{m^4}\Big{[}-8E^2(F_2m+F_3M)^2-16Em^2(F_0^2M+F_1F_2m+F_1F_3M) \nonumber \\
&+&8m^2\big{(}F_0^2(m^2+M^2)-F_1^2m^2+2F_1M(F_2m+F_3M)+(F_2m+F_3M)^2 \big{)}  \Big{]} \nonumber \\
\eeq
\end{widetext}

\section{Scalar function $I_i$}\label{sec:I_scalar}

\beq\label{eq:I_scalar}
I_1&=&\int d^3\vec{x}j_0(|\vec{p}||\vec{x}|)\delta_{\mu\nu}A_{\mu\nu}(\vec{x},t) \nonumber \\
I_{2}&=&\int d^3\vec{x}j_0(|\vec{p}||\vec{x}|)A_{00}(\vec{x},t) \nonumber \\
I_{3}&=&\int d^3\vec{x}\frac{j_1(|\vec{p}||\vec{x}|)}{|\vec{p}||\vec{x}|}x_{i}A_{0i}(\vec{x},t) \nonumber \\
I_{4}&=&\int d^3\vec{x}\frac{j_1(|\vec{p}||\vec{x}|)}{|\vec{p}||\vec{x}|}x_{i}A_{i0}(\vec{x},t) \nonumber \\
I_{5}&=&\int d^3\vec{x}\Big{\{}\frac{j_1(|\vec{p}||\vec{x}|)}{|\vec{p}||\vec{x}|}\delta_{ij}-|\vec{p}|^2\frac{j_2(|\vec{p}||\vec{x}|)}{(|\vec{p}||\vec{x}|)^2}x_ix_j}{ \Big{\}}A_{ij}(\vec{x},t) \nonumber\\
\eeq

\section{Finite-volume effects}\label{sec:FV_effect}
\begin{figure*}[htbp]
\centering
\subfigure{
\centering
\includegraphics[width=8.0cm]{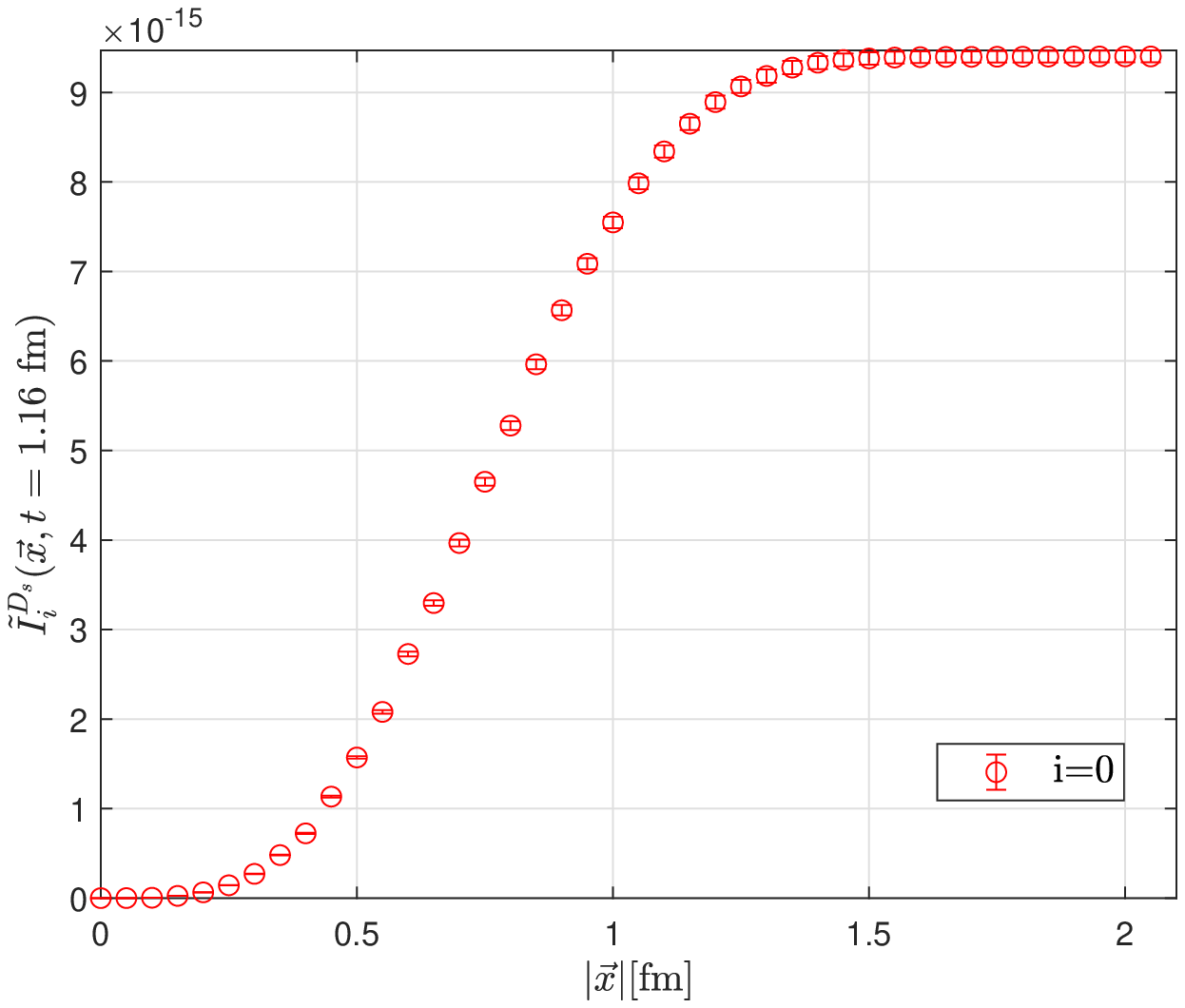}
}
\subfigure{
\centering
\includegraphics[width=8.0cm]{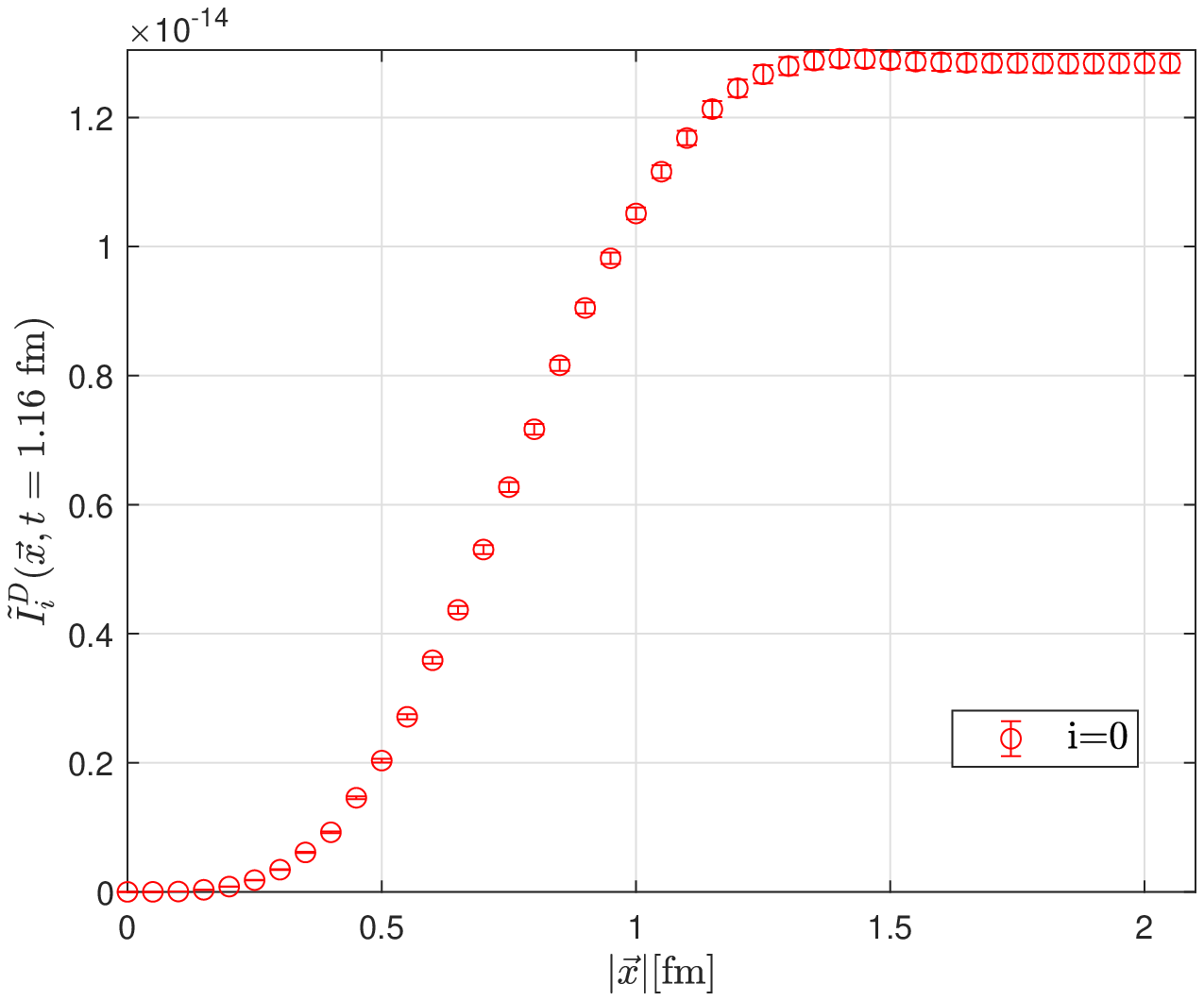}
}
\subfigure{
\centering
\includegraphics[width=8.0cm]{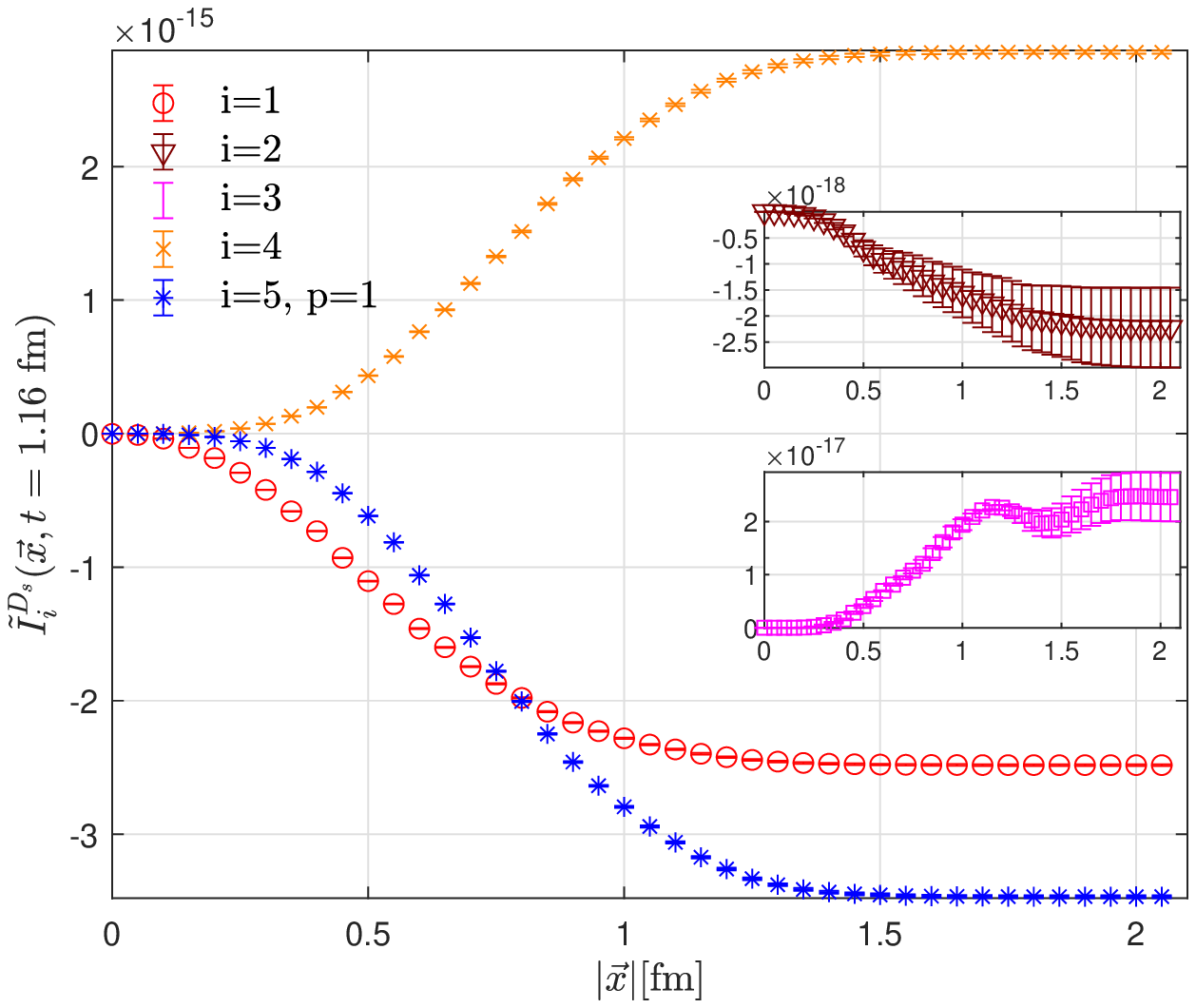}
}
\subfigure{
\centering
\includegraphics[width=8.0cm]{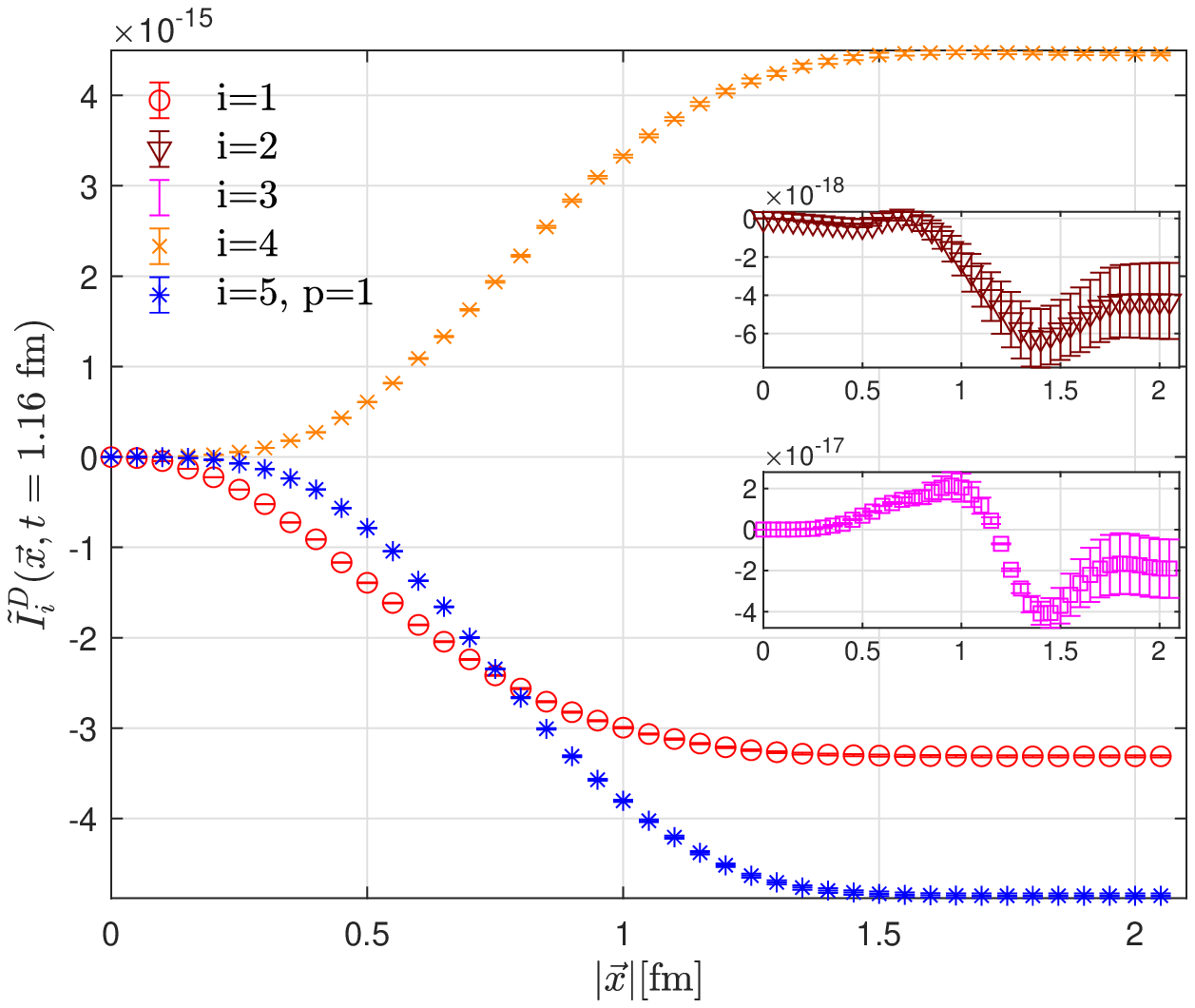}
}
\caption{\label{diag:I_V}For ensemble F32P30, $\tilde{I}_i^{D_s}(\vec{x},t)$(left) and $\tilde{I}_i^{D}(\vec{x},t)$(right) as a function of the spatial range truncation $|\vec{x}|$ for t=1.16 fm, here $i=0,1,2,3,4,5$. The p=1 denotes the projected momentum $|\vec{p}|=2\pi/L$.}
\end{figure*}

The hadronic function $H_{\mu\nu}(x)$ is dominated by the $D_s$ or $D$ state at large distance, therefore the size of the integrand for those scalar functions is exponentially suppressed when $|\vec{x}|$ becomes large. We can introduce a spatial integral truncation parameter $|\vec{x}|$ in this space integral, and study its dependence. Three volumes of the ensemble are 2.5272 fm, 2.4787 fm, 2.4898 fm for L=24, 32 and 48, respectively. The smallest is L=32. We therefore take this ensemble as an illustration. There are several structures that appear in scalar functions, as given below
\beq
\tilde{I}_0&=&\epsilon_{\mu\nu\alpha 0}x_{\alpha}V_{\mu\nu}(\vec{x},t) \nonumber \\
\tilde{I}_1&=&\delta_{ij}A_{ij}(\vec{x},t) \nonumber \\
\tilde{I}_2&=& A_{00}(\vec{x},t) \nonumber \\
\tilde{I}_3&=& x_iA_{0i}(\vec{x},t) \nonumber \\
\tilde{I}_4&=& x_iA_{i0}(\vec{x},t) \nonumber \\
\tilde{I}_5&=& \frac{j_2(|\vec{p}||\vec{x}|)}{(|\vec{p}||\vec{x}|)^2}x_ix_jA_{ij}(\vec{x},t) 
\eeq
The space integrals of these structures at a truncation parameter $|\vec{x}|$ are shown in Fig.~\ref{diag:I_V}. As far as the scalar structures in this work are concerned, there all exist plateaus for $|\vec{x}| \gtrsim 1.5$ fm, indicating that the hadronic function has negligible contribution beyond this range. Hence, the finite-volume effects should be well-controlled in our calculations. 

\section{Covariance matrix of the $q^2$-fit in Eq.~(\ref{eq:mom_extra})}\label{sec:corr_mat}
We denote the $\mathcal{C}_i^{h}(5\times 5)$ as the covariance matrix in the fitting for the form factor 
$F_i^{h}$, with $i=0,1,2,3$ and $h=D,D_s$, respectively.
\bit
\item For the ensemble C24P29 \\
\be
 \mathcal{C}_0^{D_s}=1.08\times 10^{-4}\times \left(
      \begin{array}{ccccc}
       1.00 & 0.71 & 0.51 & 0.35 & 0.26 \\
       0.71 & 0.58 & 0.45 & 0.35 & 0.27 \\
       0.51 & 0.45 & 0.42 & 0.35 & 0.29 \\
       0.35 &0.35  & 0.35 & 0.35 & 0.30 \\
       0.26 &0.27  & 0.29 & 0.30 & 0.32 \\
        \end{array}
   \right)
   \ee
\be
 \mathcal{C}_1^{D_s}=2.91\times 10^{-5}\times \left(
      \begin{array}{ccccc}
       1.00 & 0.74 & 0.58 & 0.52 & 0.48 \\
       0.74 & 0.73 & 0.66 & 0.35 & 0.56 \\
       0.58 & 0.66 & 0.73 & 0.66 & 0.60 \\
       0.52 & 0.61 & 0.66 & 0.71 & 0.66 \\
       0.48 & 0.56 & 0.60 & 0.66 & 0.74 \\
        \end{array}
   \right)
   \ee

   \be
 \mathcal{C}_2^{D_s}=1.44\times 10^{-4}\times \left(
      \begin{array}{ccccc}
       1.00 & 0.71 & 0.50 & 0.36 & 0.25 \\
       0.71 & 0.61 & 0.51 & 0.40 & 0.30 \\
       0.49 & 0.51 & 0.48 & 0.42 & 0.34 \\
       0.36 & 0.40 & 0.42 & 0.41 & 0.37 \\
       0.25 & 0.30 & 0.34 & 0.37 & 0.36 \\
        \end{array}
   \right)
   \ee

   \be
 \mathcal{C}_3^{D_s}=1.38\times 10^{-4}\times \left(
      \begin{array}{ccccc}
       1.00 & 0.70 & 0.48 & 0.32 & 0.22 \\
       0.70 & 0.57 & 0.45 & 0.34 & 0.26 \\
       0.48 & 0.45 & 0.41 & 0.35 & 0.29 \\
       0.32 & 0.34 & 0.35 & 0.34 & 0.30 \\
       0.22 & 0.26 & 0.29 & 0.30 & 0.30 \\
        \end{array}
   \right)
   \ee
\be
 \mathcal{C}_0^{D}=6.10\times 10^{-4}\times \left(
      \begin{array}{ccccc}
       1.00 & 0.62 & 0.36 & 0.21 & 0.13 \\
       0.62 & 0.47 & 0.34 & 0.23 & 0.16 \\
       0.36 & 0.34 & 0.31 & 0.26 & 0.20 \\
       0.21 & 0.23 & 0.26 & 0.26 & 0.23 \\
       0.13 & 0.16 & 0.20 & 0.23 & 0.26 \\
        \end{array}
   \right)
   \ee
\be
 \mathcal{C}_1^{D}=1.02\times 10^{-4}\times \left(
      \begin{array}{ccccc}
       1.00 & 0.54 & 0.33 & 0.29 & 0.30 \\
       0.54 & 0.59 & 0.51 & 0.41 & 0.37 \\
       0.33 & 0.51 & 0.58 & 0.54 & 0.47 \\
       0.29 & 0.41 & 0.54 & 0.63 & 0.59 \\
       0.30 & 0.37 & 0.47 & 0.59 & 0.75 \\
        \end{array}
   \right)
   \ee

   \be
 \mathcal{C}_2^{D}=8.41\times 10^{-4}\times \left(
      \begin{array}{ccccc}
       1.00 & 0.53 & 0.24 & 0.09 & 0.02 \\
       0.53 & 0.40 & 0.27 & 0.17 & 0.09 \\
       0.24 & 0.27 & 0.26 & 0.22 & 0.15 \\
       0.09 & 0.17 & 0.22 & 0.23 & 0.20 \\
       0.02 & 0.09 & 0.15 & 0.20 & 0.22 \\
        \end{array}
   \right)
   \ee

   \be
 \mathcal{C}_3^{D}=8.10\times 10^{-4}\times \left(
      \begin{array}{ccccc}
       1.00 & 0.56 & 0.29 & 0.14 & 0.07 \\
       0.56 & 0.41 & 0.29 & 0.19 & 0.13 \\
       0.29 & 0.29 & 0.27 & 0.23 & 0.19 \\
       0.14 & 0.19 & 0.23 & 0.25 & 0.23 \\
       0.07 & 0.13 & 0.19 & 0.23 & 0.26 \\
        \end{array}
   \right)
   \ee
   
\item For the ensemble F32P30
\be
\mathcal{C}_0^{D_s}=3.08\times 10^{-5}\times \left(
      \begin{array}{ccccc}
       1.00 & 0.72 & 0.52 & 0.39 & 0.29 \\
       0.72 & 0.64 & 0.51 & 0.41 & 0.33 \\
       0.52 & 0.51 & 0.50 & 0.43 & 0.37 \\
       0.39 & 0.41 & 0.43 & 0.47 & 0.41 \\
       0.29 & 0.33 & 0.37 & 0.41 & 0.45 \\
        \end{array}
   \right)
   \ee
\be
 \mathcal{C}_1^{D_s}=9.88\times 10^{-6}\times \left(
      \begin{array}{ccccc}
       1.00 & 0.71 & 0.56 & 0.43 & 0.38 \\
       0.71 & 0.71 & 0.58 & 0.54 & 0.48 \\
       0.56 & 0.58 & 0.67 & 0.57 & 0.55 \\
       0.43 & 0.54 & 0.57 & 0.68 & 0.64 \\
       0.38 & 0.47 & 0.55 & 0.64 & 0.76 \\
        \end{array}
   \right)
   \ee

   \be
 \mathcal{C}_2^{D_s}=4.87\times 10^{-5}\times \left(
      \begin{array}{ccccc}
       1.00 & 0.71 & 0.50 & 0.38 & 0.27 \\
       0.71 & 0.61 & 0.49 & 0.41 & 0.32 \\
       0.50 & 0.49 & 0.47 & 0.41 & 0.36 \\
       0.38 & 0.41 & 0.41 & 0.45 & 0.41 \\
       0.27 & 0.32 & 0.36 & 0.41 & 0.43 \\
        \end{array}
   \right)
   \ee

   \be
 \mathcal{C}_3^{D_s}=4.69\times 10^{-5}\times \left(
      \begin{array}{ccccc}
       1.00 & 0.67 & 0.46 & 0.34 & 0.24 \\
       0.67 & 0.60 & 0.47 & 0.36 & 0.28 \\
       0.46 & 0.47 & 0.44 & 0.37 & 0.31 \\
       0.34 & 0.36 & 0.37 & 0.38 & 0.33 \\
       0.24 & 0.27 & 0.31 & 0.33 & 0.35 \\
        \end{array}
   \right)
   \ee
\be
 \mathcal{C}_0^{D}=1.27\times 10^{-4}\times \left(
      \begin{array}{ccccc}
       1.00 & 0.60 & 0.35 & 0.21 & 0.14 \\
       0.60 & 0.47 & 0.33 & 0.24 & 0.19 \\
       0.35 & 0.33 & 0.32 & 0.28 & 0.21 \\
       0.21 & 0.24 & 0.28 & 0.29 & 0.25 \\
       0.14 & 0.19 & 0.21 & 0.25 & 0.30 \\
        \end{array}
   \right)
   \ee
\be
 \mathcal{C}_1^{D}=2.61\times 10^{-5}\times \left(
      \begin{array}{ccccc}
       1.00 & 0.54 & 0.32 & 0.28 & 0.29 \\
       0.54 & 0.50 & 0.41 & 0.36 & 0.33 \\
       0.32 & 0.41 & 0.46 & 0.44 & 0.39 \\
       0.28 & 0.36 & 0.44 & 0.54 & 0.51 \\
       0.29 & 0.33 & 0.39 & 0.51 & 0.72 \\
        \end{array}
   \right)
   \ee

   \be
 \mathcal{C}_2^{D}=2.00\times 10^{-4}\times \left(
      \begin{array}{ccccc}
       1.00 & 0.56 & 0.28 & 0.15 & 0.09 \\
       0.56 & 0.43 & 0.30 & 0.19 & 0.13 \\
       0.28 & 0.30 & 0.29 & 0.23 & 0.18 \\
       0.15 & 0.19 & 0.23 & 0.25 & 0.22 \\
       0.09 & 0.13 & 0.18 & 0.22 & 0.25 \\
        \end{array}
   \right)
   \ee

   \be
 \mathcal{C}_3^{D}=1.87\times 10^{-4}\times \left(
      \begin{array}{ccccc}
       1.00 & 0.58 & 0.31 & 0.17 & 0.11 \\
       0.58 & 0.44 & 0.31 & 0.22 & 0.16 \\
       0.31 & 0.31 & 0.30 & 0.25 & 0.20 \\
       0.17 & 0.22 & 0.25 & 0.27 & 0.23 \\
       0.11 & 0.16 & 0.20 & 0.23 & 0.27 \\
        \end{array}
   \right)
   \ee
\item For the ensemble H48P32
\be
\mathcal{C}_0^{D_s}=1.60\times 10^{-4}\times \left(
      \begin{array}{ccccc}
       1.00 & 0.71 & 0.51 & 0.35 & 0.25 \\
       0.71 & 0.60 & 0.50 & 0.39 & 0.30 \\
       0.51 & 0.50 & 0.51 & 0.45 & 0.36 \\
       0.35 & 0.39 & 0.45 & 0.45 & 0.37 \\
       0.25 & 0.30 & 0.36 & 0.37 & 0.39 \\
        \end{array}
   \right)
   \ee
\be
 \mathcal{C}_1^{D_s}=5.99\times 10^{-5}\times \left(
      \begin{array}{ccccc}
       1.00 & 0.71 & 0.53 & 0.42 & 0.35 \\
       0.71 & 0.68 & 0.61 & 0.51 & 0.42 \\
       0.53 & 0.61 & 0.66 & 0.59 & 0.47 \\
       0.42 & 0.51 & 0.59 & 0.61 & 0.49 \\
       0.35 & 0.42 & 0.47 & 0.49 & 0.56 \\
        \end{array}
   \right)
   \ee

   \be
 \mathcal{C}_2^{D_s}=1.88\times 10^{-4}\times \left(
      \begin{array}{ccccc}
       1.00 & 0.66 & 0.43 & 0.31 & 0.24 \\
       0.66 & 0.55 & 0.44 & 0.37 & 0.31 \\
       0.43 & 0.44 & 0.44 & 0.43 & 0.39 \\
       0.30 & 0.37 & 0.43 & 0.48 & 0.49 \\
       0.23 & 0.31 & 0.39 & 0.49 & 0.55 \\
        \end{array}
   \right)
   \ee

   \be
 \mathcal{C}_3^{D_s}=2.92\times 10^{-4}\times \left(
      \begin{array}{ccccc}
       1.00 & 0.64 & 0.45 & 0.34 & 0.28 \\
       0.64 & 0.53 & 0.45 & 0.39 & 0.33 \\
       0.46 & 0.45 & 0.45 & 0.43 & 0.38 \\
       0.34 & 0.38 & 0.43 & 0.44 & 0.41 \\
       0.28 & 0.33 & 0.38 & 0.41 & 0.42 \\
        \end{array}
   \right)
   \ee
\be
 \mathcal{C}_0^{D}=9.71\times 10^{-4}\times \left(
      \begin{array}{ccccc}
       1.00 & 0.57 & 0.29 & 0.15 & 0.11 \\
       0.57 & 0.44 & 0.32 & 0.21 & 0.14 \\
       0.29 & 0.32 & 0.32 & 0.27 & 0.20 \\
       0.15 & 0.21 & 0.27 & 0.30 & 0.24 \\
       0.11 & 0.14 & 0.20 & 0.24 & 0.30 \\
        \end{array}
   \right)
   \ee
\be
 \mathcal{C}_1^{D}=1.15\times 10^{-4}\times \left(
      \begin{array}{ccccc}
       1.00 & 0.47 & 0.26 & 0.20 & 0.15 \\
       0.47 & 0.53 & 0.48 & 0.39 & 0.29 \\
       0.26 & 0.48 & 0.55 & 0.50 & 0.39 \\
       0.20 & 0.39 & 0.50 & 0.53 & 0.46 \\
       0.15 & 0.29 & 0.39 & 0.46 & 0.59 \\
        \end{array}
   \right)
   \ee

   \be
 \mathcal{C}_2^{D}=2.30\times 10^{-3}\times \left(
      \begin{array}{ccccc}
       1.00 & 0.51 & 0.21 & 0.05 &-0.05 \\
       0.51 & 0.34 & 0.20 & 0.09 & 0.02 \\
       0.21 & 0.20 & 0.18 & 0.14 & 0.09 \\
       0.05 & 0.09 & 0.14 & 0.15 & 0.14 \\
      -0.05 & 0.02 & 0.09 & 0.14 & 0.18 \\
        \end{array}
   \right)
   \ee

   \be
 \mathcal{C}_3^{D}=6.08\times 10^{-4}\times \left(
      \begin{array}{ccccc}
       1.00 & 0.40 & 0.15 & 0.08 & 0.05 \\
       0.40 & 0.30 & 0.21 & 0.14 & 0.09 \\
       0.15 & 0.21 & 0.21 & 0.17 & 0.12 \\
       0.08 & 0.14 & 0.17 & 0.16 & 0.13 \\
       0.05 & 0.09 & 0.12 & 0.13 & 0.14 \\
        \end{array}
   \right)
   \ee
\eit

\end{document}